\documentclass[11pt]{article}
\usepackage[margin=1.2in]{geometry}
\usepackage[T1]{fontenc}
\usepackage[utf8]{inputenc}
\usepackage{microtype}
\usepackage{graphicx}
\usepackage{float}
\usepackage{placeins}
\usepackage{caption}
\usepackage{array}

\usepackage[hidelinks]{hyperref}
\hypersetup{pdftitle={INDRA: A New AI Tool for Exploring Tobacco, Fossil Fuel, and Chemical Industry Archives},pdfauthor={Daniel Akselrad and Robert N. Proctor}}
\usepackage{xurl}
\usepackage{enotez}
\setenotez{list-name=Notes, totoc=false}
\begin{document}

\vspace*{-51pt}
\begin{center}
{\LARGE INDRA: A New AI Tool for Exploring\\[2pt] Tobacco, Fossil Fuel, and Chemical Industry Archives\par}
\vspace{16pt}
{\large Daniel Akselrad$^{1}$ and Robert N. Proctor$^{2}$\par}
\vspace{11pt}
{\small $^{1}$Department of Epidemiology \& Population Health, Stanford University, Stanford, CA, USA\\
$^{2}$Department of History, Stanford University, Stanford, CA, USA\par}
\end{center}
\vspace{-8pt}

\begin{center}\textbf{Abstract}\end{center}
\begin{quotation}\small\noindent
{\looseness=-1 Five decades of litigation have disgorged hundreds of millions of pages of formerly secret business records from the tobacco industry, along with documents from the makers of drugs, chemicals, food, firearms, and fossil fuels. Yet these archives have been effectively inaccessible to general-purpose large language models (LLMs) because they have never been compiled into an LLM-readable corpus. Chatbots may be familiar with some of the materials contained in such archives but, with no direct access to the documents, they are vulnerable to hallucination and other defects. Here we introduce INDRA, a research platform designed to remedy such failures by embedding the conventions of archival historiography into a system-level protocol governing every output. The platform federates UCSF’s Industry Documents Library, Columbia and CUNY’s ToxicDocs, Stanford’s SRITA, and other heretofore siloed collections, and provides three interlinked safeguards: (1) a closed evidentiary sandbox confines the model to a user-selected corpus, blocking retrieval from external sources that could introduce bias; (2) real-time provenance tagging marks the boundary between archival evidence and parametric inference; and (3) a system-level protocol enforced by deterministic scripts guides the structure of every output. Together these safeguards prevent the model from conflating “the documents say X” with “I think X” or “I learned X from prior training.” The result is an LLM-powered research partner enabling massive multi-archival investigations, a tool whose outputs are designed to be checked rather than trusted, and whose architecture makes the conditions of knowledge production visible and auditable. Three case studies demonstrate the method’s analytical value and limitations, including what we call the \emph{Heraclitus effect}, the \emph{steppingstone dilemma}, and the \emph{gullibility }(or\emph{ mafia})\emph{ problem}.\par}
\end{quotation}
\vspace{-10pt}

\FloatBarrier
\section{Introduction \& Background}

{\looseness=-1 \enlargethispage{2\baselineskip}Five decades of U.S. litigation have disgorged hundreds of millions of pages of formerly secret business records from the tobacco industry, along with documents from the makers of drugs, chemicals, food, firearms, and fossil fuels.\endnote{Stanton A. Glantz et al., “Looking through a Keyhole at the Tobacco Industry: The Brown and Williamson Documents,” \emph{JAMA} 274, no. 3 (July 19, 1995): 219–24; Stanton A. Glantz, John Slade, Lisa A. Bero, Peter Hanauer, and Deborah E. Barnes, \emph{The Cigarette Papers} (Berkeley: University of California Press, 1996); Ruth E. Malone and Edith D. Balbach, “Tobacco Industry Documents: Treasure Trove or Quagmire?” \emph{Tobacco Control} 9, no. 3 (2000): 334–38; David Rosner, Gerald Markowitz, and Merlin Chowkwanyun, “ToxicDocs (www.ToxicDocs.org): From History Buried in Stacks of Paper to Open, Searchable Archives Online,” \emph{Journal of Public Health Policy} 39, no. 1 (2018): 4–11. Litigation has been the primary source of such documents, but the original prompt for such litigation has often been courageous acts of whistleblowing, FOIA requests, and occasional leaks. Brown \& Williamson’s 1969 “Doubt is our product” memo, for example, was leaked from the Federal Trade Commission in 1981.} The UCSF Industry Documents Library, for example, contains some 150 million pages that would take even a quick reader thousands of years to digest. Techniques to retrieve and analyze these records have evolved over time yet still suffer significant limitations. Keyword search and topic modeling can retrieve but not interpret, qualitative coding interprets but cannot scale, and feeding all 400 billion “tokens” of text into a general-purpose AI model would currently cost millions of dollars in user fees. Large language models (LLMs) also often hallucinate, blurring the line between verifiable facts and untethered confabulations.\endnote{Adam Tauman Kalai, Ofir Nachum, Santosh S. Vempala, and Edwin Zhang, “Why Language Models Hallucinate,” arXiv preprint arXiv:2509.04664, September 4, 2025, \url{https://arxiv.org/abs/2509.04664}. One reason models hallucinate is that pre-training rewards replicating training data rather than validating truth. Because benchmarks penalize “I don’t know” the same as a wrong answer, models are incentivized to make an educated guess.}\par}

{\looseness=-1 How can we explore hundred-billion-word archives with epistemic integrity and at scale? And how can we extend this kind of analysis to the next mega-tranche of archived documents nominally public, yet practically inaccessible due to clumsy or obsolete interfaces (sometimes deliberately so)?\endnote{The R. J. Reynolds litigation-document website (rjrtdocs.com) required by the 1998 Master Settlement Agreement and the federal RICO judgment in \emph{United States v. Philip Morris USA}, 449 F. Supp. 2d 1 (D.D.C. 2006) has been effectively frozen since September 2021, the endpoint of the judgment’s fifteen-year continued-disclosure remedy. The site’s corpus added 160,993 documents in 2000 but zero in 2023 and 2024, and the user interface (built on the DocuBase platform using antiquated JavaScript) has remained largely unchanged since its launch. Much like a hospital that doesn’t want patients or a school hostile to learning, such adversarial archives are designed to be difficult to use.} Finally, how can we move beyond some of the most notorious limitations of LLMs: sycophancy, hallucination, “lost in the middle” degradation, the “reversal curse,” and the myth of the wisdom of crowds on the open internet?\endnote{Mrinank Sharma et al., “Towards Understanding Sycophancy in Language Models” (Anthropic, 2023), arXiv:2310.13548; Ziwei Ji et al., “Survey of Hallucination in Natural Language Generation,” \emph{ACM Computing Surveys} 55, no. 12 (2023): 1–38; Nelson F. Liu, Kevin Lin, John Hewitt, Ashwin Paranjape, Michele Bevilacqua, Fabio Petroni, and Percy Liang, “Lost in the Middle: How Language Models Use Long Contexts,” \emph{Transactions of the Association for Computational Linguistics} 12 (2024): 157–73; Lukas Berglund, Meg Tong, Max Kaufmann, Mikita Balesni, Asa Cooper Stickland, Tomasz Korbak, and Owain Evans, “The Reversal Curse: LLMs Trained on ‘A is B’ Fail to Learn ‘B is A,’” in \emph{Proceedings of the Twelfth International Conference on Learning Representations} (ICLR, 2024). Sycophancy refers to the tendency of LLMs to be overly agreeable or even obsequious toward users. This occurs because models are designed to learn from human feedback.} \par}

Here we introduce a new tool to access, analyze, and interpret massive online archives in ways that protect against many of the traditional dangers of research using LLMs. INDRA (for Industry Documents Research and Analysis) is an AI-powered platform built on top of Anthropic’s Claude language model, designed to help researchers interrogate vast and ever-expanding archives of corporate documents. INDRA allows anyone with an internet connection to search across UCSF’s Industry Documents Library, Columbia University and CUNY’s ToxicDocs, Stanford’s SRITA, and other now-federated archives, and to enter into an evidence-based conversation exploring a targeted set of documents.\endnote{The federated archives and their holding institutions: Industry Documents Library, University of California, San Francisco, \url{https://www.industrydocuments.ucsf.edu/}; ToxicDocs, Center for the History and Ethics of Public Health, Columbia University, and the Graduate Center, City University of New York, \url{https://www.toxicdocs.org/}; Stanford Research into the Impact of Tobacco Advertising (SRITA), Stanford University School of Medicine, \url{https://tobacco.stanford.edu/}; Trinkets \& Trash, Rutgers School of Public Health, \url{https://www.trinketsandtrash.org/}; FBarchive, Public Interest Tech Lab, Harvard Kennedy School, \url{https://fbarchive.org/}; the Oxford Carbon and Climate Advertising Library (OxCCAL), Climate Litigation Lab, Oxford Sustainable Law Programme, University of Oxford, \url{https://www.smithschool.ox.ac.uk/research/climate-litigation-lab}; and the R. J. Reynolds litigation document website, \url{https://www.rjrtdocs.com/}. INDRA additionally holds its own collections. This list will continue to expand.} Combining these heretofore siloed collections enables intelligent cross-corpus pattern recognition with a speed and efficiency not previously possible. Researchers wanting to explore networks, strategies, and rhetorics common to the makers of cigarettes, plastics, and fossil fuels, for example, no longer need to leap from one website to another but can access either all of these at once or a deliberate selection, through a single portal.

This new platform incorporates three interlinked safeguards. First and foremost, a closed \emph{evidentiary sandbox} confines the model to a user-selected corpus, blocking retrieval from external sources that could introduce bias while capping working memory below the point at which LLM performance starts to degrade. Second, real-time \emph{provenance tagging} distinguishes archival evidence from the model’s prior learning. And third, to strengthen the model’s fidelity to the archives, a system-level protocol comprising thirty instructions governs every output. This protocol is backed by pre-filled assistant turns and a deterministic “watchdog” script (involving no LLMs) that checks each quotation against the original source OCR.\endnote{The model’s response is held until the quotation watchdog is finished scanning for fabrications; any detected hallucinations are removed before they can ever be read by the user.} Together, these safeguards enforce citational discipline, verbatim extraction, hallucination prevention, and disclosure of negative findings (it will tell you if something is \emph{not} in the documents selected for analysis). Preliminary testing shows that when provided with the same documents, INDRA’s combined safeguards significantly reduce rates of fabrication and misgrounding when compared with Anthropic’s off-the-shelf Claude Sonnet 5. And while the bare model with no safeguards responds confidently with wrong answers under increased document loads, INDRA refuses to provide an output of which it is uncertain.\endnote{During testing, when Sonnet 5 was provided with no documents at all (as is often the case when using off-the-shelf LLMs to ask about these archives), 65 percent of its responses contained unverifiable quotations, and not one of them offered a resolvable document ID. When we loaded both INDRA and Sonnet 5 with identical document sets, INDRA fabricated no quotations (0 of 3,643 across both loads), while the bare model, Sonnet 5, delivered a handful of quotation-marked passages found in no loaded documents. The bare model misgrounded real quotes (crediting them to the wrong document) 5.1 percent of the time under light loads and 9.2 percent under heavy loads. INDRA’s rate held at 0.5 and 0.6 percent. See Appendix A.1, “Integrity.” Appendices are available at \url{https://indra.stanford.edu/methods/appendices/}. Appendices hereafter are cited by letter, section, and title. On attribution with LLMs, see Hannah Rashkin et al., “Measuring Attribution in Natural Language Generation Models,” \emph{Computational Linguistics} 49, no. 4 (2023): 777–840.}

Working in this way allows INDRA to perform rapid analysis and interpretation of industry tactics and rhetorics, for example, and how these have evolved over time. Networks of corporate collaboration can be explored, along with subtle aspects of workplace culture and connivance. One can inquire into the most common or vivid metaphors used in a particular universe of documents, or even look for metonymy or synecdoche, or discover what kinds of words or ideas cluster with one another in cross-industry communications. Near-instantly one can reconstruct chronologies, typologies, meaning and intent, evidence of absence, and countless other aspects of the documents contained in these archives—all of which can be generated in more than a dozen different languages.

Building such a tool poses several methodological challenges, starting with the vastness of these archives. The Industry Documents Library alone contains hundreds of billions of tokens of text (a token is roughly two characters), the equivalent of millions of scholarly books.\endnote{Different models use different tokenizers, which means that token counts are not a fixed property of the text. The same document can be, say, 120,000 tokens for one model yet 140,000 for another. INDRA applies a generous conversion ratio (roughly 1.8 characters per token) to prevent the token budgeting gate from admitting too many documents.} Traditionally, a researcher asking “how did company X first internally acknowledge risk Y?” must sift through duplicates and learn to ignore countless stray and irrelevant dead ends and cul-de-sacs. An additional problem even with computer-assisted keyword or “snowball searches” is that such searches are hyper-literal: they find but cannot organize, interpret, or compare key terms, including fuzzy matches across sources.\endnote{On snowball sampling, see Lisa Bero, “Implications of the Tobacco Industry Documents for Public Health and Policy,” \emph{Annual Review of Public Health} 24, no. 1 (2003): 267–88.} To obtain a thorough list of what terms cigarette makers use to denigrate public health advocates (besides “anti-smokers,” “fumeophobes,” or “the tobacco Taliban”), for example, requires an intelligent understanding of the meaning and intent of such terms in their historical context.\endnote{The industry’s list of pejoratives includes \emph{zealots}, \emph{militants}, \emph{fanatics}, and \emph{antis}, but also \emph{misocapnists}, \emph{control freaks}, and \emph{shower adjusters}.} Franco Moretti identified this predicament three decades ago, but his method of “distant reading” (network analysis, topic modeling, quantitative stylistics) was designed for literary texts whose authors had no reason to believe they would be brought into court for fraud, conspiracy, or negligence.\endnote{Franco Moretti, \emph{Distant Reading} (London: Verso, 2013); Ted Underwood, \emph{Distant Horizons: Digital Evidence and Literary Change} (Chicago: University of Chicago Press, 2019).} Corporate documents produced through litigation, however, are different in kind. They are often strategically constructed and sometimes even deliberately misleading; these are \emph{adversarial archives}, meaning collections divulged under duress or by court order, and which therefore cannot always be taken at face value.\endnote{Ann Laura Stoler, \emph{Along the Archival Grain: Epistemic Anxieties and Colonial Common Sense} (Princeton: Princeton University Press, 2009); Michel-Rolph Trouillot, \emph{Silencing the Past: Power and the Production of History} (Boston: Beacon Press, 1995).} Topic modeling can identify clusters of co-located terms, but it cannot distinguish between an honest internal assessment and a memo written with one eye on future discovery—a type of rhetoric we call \emph{eavescasting}.\endnote{Daniel Akselrad, Pamela M. Ling, Tracey J. Woodruff, Nicholas Chartres, and Robert N. Proctor, “Tobacco Industry Litigation Strategies: How Cigarette Attorneys Manufacture Doubt in Court” (under review); Trial Transcript, vol. 19, testimony of Robert N. Proctor, 2547, \emph{Campbell v. R.J. Reynolds Tobacco Co.}, No. 2011-CA-005960 (Fla. Cir. Ct., 10th Jud. Cir., Polk Cnty., May 15, 2013), UCSF Industry Documents Library (UCSF IDL), \url{https://www.industrydocuments.ucsf.edu/docs/ksbn0225/}.}

General-purpose LLMs such as ChatGPT, Gemini, Claude, and even DeepSeek introduce an entirely new set of limitations. When a researcher uses these off-the-shelf tools to conduct historical research, the model is typically unable to deliver unfettered access to archived documents, whether text, image, or video. Such models are often “sandboxed” but in an inverted way: Anthropic et al. would like nothing more than to be able to scrape or ingest \emph{all} knowledge (for parametric learning), yet as of now they are largely barred from providing users with direct access to books under copyright, archival documents, and published scholarly papers.\endnote{For the legal case distinguishing AI training from reproduction or distribution to users, see \emph{Bartz et al. v. Anthropic PBC}, No. 3:24-cv-05417 (N.D. Cal. June 23, 2025), in which Judge William Alsup ruled that training on lawfully acquired books constituted fair use but that retention of pirated copies did not. Anthropic subsequently acquiesced to a \$1.5 billion class-action settlement covering approximately 500,000 pirated works.} Many archives, even newspapers, are nominally “public” yet exist behind paywalls, or block AI traffic to prevent their data from being scraped or even fetched. Like most historical archives, the Industry Documents Library is inaccessible to AI chatbots like Claude or ChatGPT; INDRA gets around this by federating the collections into an AI-accessible form. (This was no simple task: ingesting all 150 million pages into a Cloudflare R2 bucket alone took six weeks of round-the-clock processing.)

An altogether different problem with general-purpose LLMs is that they almost always allow users to max out or even exceed the chatbot’s working memory (its context window) or to engage in overly long conversations, discarding material once some limit is reached. This degrades recall and increases the risk of hallucination by causing “lost in the middle” effects or “attention sink,” whereby models focus more on how a document (or AI chat) begins and ends than on what is in the middle.\endnote{Transformers exhibit a U-shaped attentional bias: they weight material located at the beginnings and ends of long texts more heavily. There are a few probable reasons for this, though the causes are uncertain. One explanation is that documents used for training concentrate valuable framing material in start and end locations such as titles, prefaces, introductions, and conclusions. Another is that models are designed to predict each word based on the words that come before it, and the opening words always come before later text. Humans, too, can exhibit such a primacy or recency bias depending on the situation. See Liu et al., “Lost in the Middle”; Guangxuan Xiao, Yuandong Tian, Beidi Chen, Song Han, and Mike Lewis, “Efficient Streaming Language Models with Attention Sinks,” in \emph{Proceedings of the Twelfth} \emph{International Conference on Learning Representations} (ICLR, 2024).} When the context window is filled to whatever maximum token limit is allowed, the model will proceed as if the question can be answered from an impoverished abridgment, and fills gaps with plausible-sounding fabrications based on its attentional bias toward the beginnings and ends of documents or document sets. The result is that LLMs are often biased in how they will read a document and become prone to referencing real-seeming yet non-existent documents. They may even fabricate quotations or document ID numbers not from malice, but because they are optimizing for fluency—and sycophancy—rather than veracity. The output might sound like scholarship but is often epistemologically hollow, lacking attribution and factual grounding. For legal discovery, historical research, and regulatory review, an unprovable inference can be perilous for scholarship or a court case, or even for human or planetary health.\endnote{Varun Magesh, Faiz Surani, Matthew Dahl, Mirac Suzgun, Christopher D. Manning, and Daniel E. Ho, “Hallucination-Free? Assessing the Reliability of Leading AI Legal Research Tools,” \emph{Journal of Empirical Legal Studies} 22, no. 2 (2025): 216–42 (finding hallucination rates between 17 and 33 percent despite vendor claims); \emph{Kohls v. Ellison}, No. 24-cv-3754 (D. Minn. Jan. 10, 2025), in which Judge Laura M. Provinzino excluded a Stanford expert’s declaration on AI misinformation after determining that the declaration itself contained fabricated citations generated by ChatGPT, “shatter[ing]” the expert’s credibility before the court.}

Scholars have attempted to address some of these challenges, albeit imperfectly. Retrieval-augmented generation (RAG) systems,\endnote{Patrick Lewis et al., “Retrieval-Augmented Generation for Knowledge-Intensive NLP Tasks,” in \emph{Advances in Neural Information Processing Systems} 33 (2020): 9459–74. A RAG system is an AI architecture designed to provide responses based on passages retrieved from documents loaded into context rather than from parametric memory. See also Noah J. Kim-Baumann and Torsten Hiltmann, “HistoRAG: Embedding Historical Methodology in Retrieval-Augmented Generation Through Critical Technical Practice,” arXiv preprint arXiv:2606.18103 (2026).} the standard architecture for “chat with your documents” apps, extract relevant passages but may lack the evidentiary tiering, provenance tagging, and hallucination prevention protocols required for critical and responsible historiography. Some pioneering tools, such as Oxford’s Corporate Litigation and Accountability Research Assistant (CLARA) or KinoAI’s Jmail/Jemini, offer document analysis with citation linking but do not presently distinguish between domain knowledge and archival data.\endnote{Meghana Patakota, Jake Rutherford, and Benjamin Franta, Corporate Litigation and Accountability Research Assistant (CLARA), Climate Litigation Lab, Oxford Sustainable Law Programme, University of Oxford, 2025, accessed August 1, 2026, \url{https://clara-research.com}; Luke Igel and Riley Walz, \emph{Jmail}, Kino AI, 2025, accessed August 1, 2026, \url{https://jmail.world}.} They do not flag interpretation, and they do not provide adequate safeguards against untraceable confabulation or “lost in the middle” degradation (a.k.a. U-shaped attention bias). What remains unaddressed in these platforms is the dilemma facing any computational archival investigation: (1) how to make visible the crucial difference between what can be proven from the archives and what the machine “believes,” and (2) how to operationalize the earnest skepticism required by historical inquiry, while simultaneously educating users on how LLMs read, “think,” and respond along the way.

{\looseness=-1 INDRA differs from these approaches because it incorporates the epistemic conventions of archival historiography into a system-level protocol that governs every output. Its design logic is shared with RAG systems, but where prior systems address provenance simply by delivering citations, INDRA makes visible how the model is drawing its conclusions from the material. Where a standard RAG pipeline retrieves passages and lets the model speak freely about them, INDRA inserts checkpoints between retrieval and generation, and will not answer until all gates are cleared: a tier label, but also a provenance tag, an OCR quality assessment, verified exact quotations, and reporting on the absence of evidence when evidence is absent. While commercial AI suppliers advertise million-token context windows and limitless conversations as features, the present method regards both as liabilities for researchers who value accuracy and citational integrity. INDRA places a hard limit on how many tokens the model is allowed to hold in memory during any one session, and terminates sessions before “lost in the middle” effects might otherwise corrupt recall and well before any context truncation can occur. \par}

{\looseness=-1 Unlike prior designs, where relevance scoring decides which documents the model will see, this new way of working places this decision entirely in the hands of the researcher. Documents enter the sandbox only by deliberate selection, and the model cannot retrieve or even consider any materials mid-session that have not already been selected for analysis. This avoids the paternalism of previous designs, where the model decides which documents are relevant by ranking them solely against pre-digested vector-embeddings over which the user has no control. Our design, by contrast, enables exploration of winding side streets that exist beyond the broad avenues any single search method can map. And since even a rich graph can never map the entirety of the federated archives, INDRA enables both semantic and Boolean search, allowing investigation of subtle or coded phrases such as “our friends” or “please destroy” that no embedding or entity extractor would have recognized as important.\par}

\enlargethispage{2\baselineskip}{\looseness=-1 Put another way, the fundamental question of quality control (and utility!) for bringing LLMs to bear on such archives is not just whether any given citation is accurate; documents could be misread or misquoted even if properly identified. More importantly, it is whether the platform makes it possible for researchers (a) to find the documents needed to answer their questions and (b) to distinguish conclusions derived from the archive from those based on the model’s prior training (i.e., domain knowledge). The aim is not to usurp or supplant the scholar, but rather to equip them with an intelligent interlocutor for archival corpora, a tool whose outputs are designed to be verified, not trusted, and whose architecture makes the conditions of knowledge production more visible and auditable. We demonstrate this new platform through three case studies, each illustrating the advantages of working with an LLM harnessed to allow simultaneous exploration of multiple archives from different industries. We also examine the limitations that any such design must confront, including the \emph{Heraclitus effect}, the \emph{steppingstone dilemma}, and the \emph{gullibility }(or\emph{ mafia})\emph{ problem}, among others.\par}

\FloatBarrier
\section{Methods}

\FloatBarrier
\subsection{The Evidentiary Sandbox}

{\looseness=-1 INDRA is built on a foundational principle borrowed from archival science: researchers should be able to interrogate documents in ways that are maximally faithful to the meaning (or plain text) of the original, with steadfast attention to provenance. Central to this architecture is a closed \emph{evidentiary sandbox}, which applies a now-standard pattern in RAG systems to public health research, where it has been conspicuously absent.\endnote{Closed-corpus RAG design is today becoming standard across commercial legal, financial, and enterprise AI tools. In these systems, a general-purpose model engages with a specific set of documents and returns answers linked to them. See, e.g., Harvey AI (in law) and Hebbia (in finance). Harvey (Counsel AI Corporation, San Francisco), accessed August 1, 2026, \url{https://harvey.ai}; Matrix (Hebbia, New York), accessed August 1, 2026, \url{https://hebbia.com}.} Here the language model receives only the OCR-extracted text of user-selected documents and is explicitly forbidden from drawing on outside sources, apart from whatever may have originally been acquired from prior training (i.e., semantics and basic logic but also domain knowledge—which is flagged in any output). The archives to which INDRA has access constitute the entirety of the model’s evidentiary base: no retrieval from external databases is allowed, and access to the open web is blocked by design. Whereas platforms such as Google’s Gemini Notebook will sporadically search the web to fill gaps in their knowledge, our closed-context architecture does not allow access to URLs, APIs, or any other source beyond the OCR text it has been given in the form of JSON files.\endnote{The closed/open distinction in LLM-generated content has roots in the “closed-book” vs. “open-book” framing developed in Adam Roberts, Colin Raffel, and Noam Shazeer, “How Much Knowledge Can You Pack into the Parameters of a Language Model?,” in \emph{Proceedings of the 2020 Conference on Empirical Methods in Natural Language Processing }(Association for Computational Linguistics, 2020), 5418–26.} The protocol requires every factual claim to be traceable to a specific document in the chosen set, with document IDs provided for independent checking, or explicitly flagged as drawing on knowledge from pretraining. To ensure parity with the underlying archives, INDRA audits itself against its own sources nightly. To reduce response variability, the model is set to run at temperature zero where available; and where models no longer support the temperature parameter, deterministic scripts keep the LLM grounded. This makes it possible to increase the consistency of outputs and to prevent speculation beyond the provided corpus.\endnote{INDRA sets temperature=0 wherever models still accept it; this is the setting that holds a model to its most probable wording. Where the Anthropic API has retired temperature in favor of “extended thinking” (beginning with Sonnet 5), INDRA switches this off and relies on the quote-anchoring watchdog to catch and kill fabrication before it can reach the user. This addresses the same grounding concern that the temperature=0 parameter guards against.} When asked “Who is Rosa Parks?” for example, if she is not identified in the selected documents it will answer, “I don’t know” or “I cannot answer this based on the documents provided,” rather than filling in the gap from memory. Taken together, the LLM’s role is akin to that of a research assistant confined to a reading room: it can analyze, organize, and compare what is given to it on the table, but it cannot check the shelves or leave the room.\par}

{\looseness=-1 This raises a methodological question: How can sandboxing of this sort mitigate the problem of hallucination and ensure reliable, verifiable outputs during analysis? One way is to place a hard ceiling on INDRA’s ability to fill the LLM context window. Liu et al., for example, identify a recall drop once the model’s working memory is filled, and recommend two design fixes to reduce the risk of hallucination: (1) re-ranking retrieved documents so that the most relevant evidence occupies the head of the prompt, and (2) truncating retrieval rather than overwhelming it. INDRA implements both of these by capping per-query document retrieval below an empirically derived 50 percent threshold,\endnote{During testing, we observed that the model’s quote accuracy without INDRA’s safeguards in place fell from 96 percent to 61 percent as the context window filled to a million tokens. At its 500,000-token limit, every quotation INDRA retrieved was exact and correctly attributed. When it could not verify a sentence, it declined rather than guessed, preventing fabrication. See Appendix A.2, “Context Utilization.”} and by ordering documents by relevance when they enter the prompt. We add to this a third constraint: a ceiling placed on conversation length. INDRA tells users how much of their context budget (and the model’s hard cap) they have used in any given session by providing a real-time accounting of token usage. By enforcing a 500,000-token context limit for documents and leaving 50,000 tokens of headroom for system prompts, pre-filled assistant turns, and the user’s conversations, we find the model produces more reliable results.\endnote{Respecting this hard ceiling means that any single document greater than 500,000 tokens in size cannot currently be analyzed by the tool directly (these are rare, under half of one percent of the federated archives). For these documents too large to be analyzed directly, INDRA slices them into chunks of $\sim$100,000 tokens each, choosing an intelligent point in the document in which to make the slice (page boundaries, row or sheet boundaries for spreadsheets, etc.). Slices, made deterministically with no LLMs, include an overlap so that any passage crossing a page boundary can be reconstructed and verified by the quotation watchdog.}\par}

{\looseness=-1 As for the “reversal curse”: in 2023, Berglund et al. discovered that LLMs trained on a fact in one direction (“Olaf Scholz was the ninth chancellor of Germany”) often fail to generalize when asked in the reverse (“Who was the ninth chancellor of Germany?”).\endnote{Berglund et al., “The Reversal Curse.”} Such retrieval asymmetries are baked into the model’s weights at training time and cannot be prompted away. The reversal curse as such is a property of parametric memory, and INDRA’s architecture largely removes the model’s reliance on this type of retrieval: the archival evidence sits in the context window rather than in the weights so that a fact stated in one direction can be correctly read in the reverse. Where the model does draw on its training (and it inevitably does), that contribution is marked in-line rather than blended with the archival record (see also below). Quotations at analysis time are retrieved from the documents’ actual text rather than from vector embeddings, so relational structure is never compressed into a geometry that might distort it.\par}

\FloatBarrier
\subsection{Provenance Tagging}

{\looseness=-1 Novel in this system is a measure designed to indicate to the user how the model is sourcing its assertions at the sentence level. Provenance tagging goes beyond traditional in-line citation in that INDRA reveals where and how it has derived its output, along with a metric of the integrity (i.e., OCR quality) of the underlying material. When the model draws from outside the provided documents, it will flag the claim with one or more of three distinct markers: \textbf{Domain Knowledge }[DK] for claims relying on its background factual knowledge or training; \textbf{Interpretation }[INT] for claims that make inferences beyond what could be obtained through a literal reading of the documents; and/or \textbf{Perspective Shift }[PS] for passages written from a user-requested perspective (e.g., explain this document from the perspective of a fossil fuel executive, or a born-again Christian, or a gun rights activist, or even an endangered coral reef). These provenance tags map nicely onto scholarly norms of close reading, where a researcher strives to identify where an author might be coming from politically, ethically, or ontologically. Perspective shifting can be useful for stress-testing a finding: when a researcher requests an analysis from a specific point of view, the system commits to that lens, with every perspective-driven sentence tagged to distinguish “painted” claims from a more neutral report (see Figure 1).\par}

{\looseness=-1  Flagging conclusions derived solely from domain knowledge becomes especially important when the model attempts to rely on a finding or document from its training that is real and genuine—it’s in the archives—but has not been selected for the sandbox. This happens because many of the industry’s documents are often written about in newspapers and in public litigation records (or cited in scholarship) that the model may know about from its training. INDRA correctly regards these texts recalled from training as of questionable reliability and tags them as domain knowledge—because LLMs are more likely to confabulate quotations from documents they hold in parametric memory than in their context window.\endnote{When this occurs, these sources are not recalled from cached memory, the model has not escaped the sandbox, and there is no backdoor. In early work with INDRA, the underlying model recalled document ID ngvm0216, a real 1971 letter by W.T. Hoyt in the CTR Records located in the UCSF IDL, and the model flagged it as deriving from domain knowledge [DK]. This occurred because the IDL is public, and CTR Special Projects documents are among the most litigated and written-about documents anywhere. Their IDs and contents are in Anthropic’s training corpus, so when reasoning about Hoyt’s relationship to CTR Special Projects and SP-123 funding, the model completed the pattern with a real, relevant ID it had memorized at training time. INDRA is instructed never to rely on quotations from these documents recalled from parametric memory because such quotations are more likely to contain inaccuracies.} Warning that these inputs derive from domain knowledge protects researchers from relying on claims the model cannot back up with direct quotations from the corpus. Importantly, the system does not rely on the model’s self-reporting to judge which claims draw on domain knowledge: a deterministic watchdog script identifies and tags any claim that is not verifiably present in the context window so that [DK] markers indicate an external grounding rather than the model’s own self-report. Open-weight models may make it easier to tell whether any given fact or phrase derives solely from parametric memory; we are currently experimenting with this.\par}

\begin{figure}[p]
\centering
\fbox{\includegraphics[width=1.00\textwidth,height=0.81\textheight,keepaspectratio]{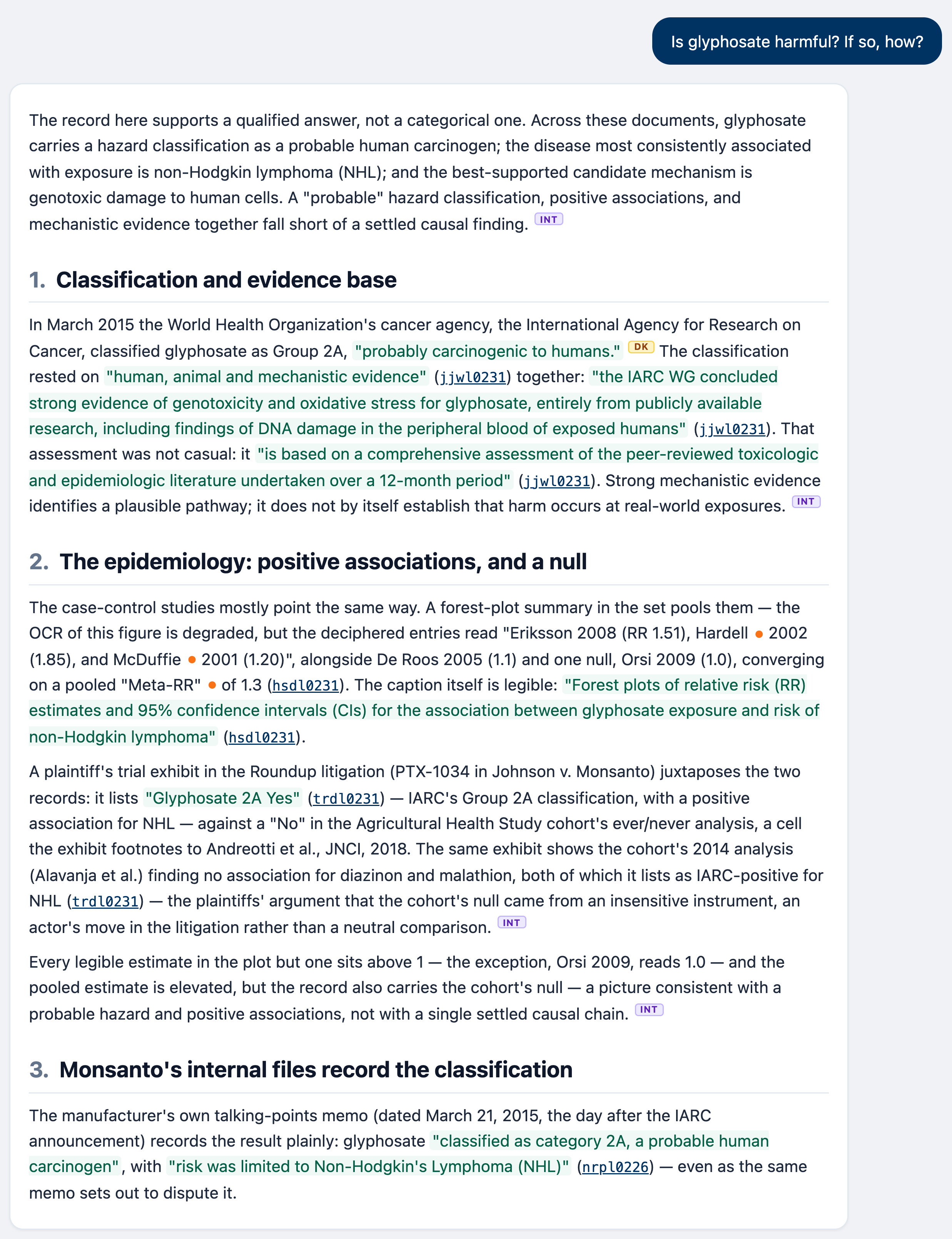}}
\caption{INDRA session with provenance tags. INDRA notifies the user when it has drawn conclusions that are clearly interpretive (i.e., moving beyond direct quotation to complex inference or evidence-based opinion) or based on the model’s broader domain knowledge; it knows about the 1964 Surgeon General’s report, for example, but will not reference this without labeling it as [DK]. When asked to write from a certain point of view, INDRA labels its responses with a “Perspective Shift” tag. When its confidence in the quality of the OCR on any given document is low, an OCR quality warning is issued as an in-line tag, here in orange.}
\end{figure}

Establishing provenance matters all the more when different archives—and different documents within the same archive—provide inputs of varying legibility. When INDRA fetches documents for analysis, it does not ingest the PDF files housed in the archives but instead draws on optical character recognition (OCR) files as proxies for the original documents. But even the most carefully scanned files can contain errors due to mechanical misreadings, especially in older documents from the carbon copy era or handwritten notes. This is particularly important given that readability and significance are often inversely correlated, because damning documents that are hard to read or in foreign text (\emph{Oświęcim} vs. \emph{Auschwitz}) seem to have more easily escaped the vigilance of industry attorneys responding to subpoenas.\endnote{Daniel Akselrad and Robert N. Proctor, “Why Did Philip Morris Stop Making Cigarettes at Auschwitz? An Essay on the Geometry and Kinetics of Atrocity,” \emph{Public Culture} 36, no. 1 (2024): 47–74.} Files added to the archives fifteen years ago were also limited by the OCR engines available at that time. Newer engines can still make mistakes: some produce reliable transcriptions in some contexts but not in others. So there is a high degree of variance in OCR quality. To remedy this, archival platforms must have a valid method for scoring and displaying OCR confidence so that users will know whether they will need to examine the document with their own eyes (assuming the user is not an AI agent).\endnote{Because no human-verified transcriptions of these documents exist to cross-check against, OCR quality is estimated from the recovered text alone. Each document is classified by type so that a page of numbers is not misjudged to be poorly-constructed natural language. We then score the document for signs of misrecognition (nonsense characters, letter–digit errors such as “h3llo,” column-induced word breaks, etc.) and type-appropriate recognizability (dictionary words for prose; plausible dates and units for tables). The result is a confidence score between 0 and 100. This method screens for legibility, not fidelity, so a misread of one real word for another can still pass.} INDRA thus provides an in-line OCR quality warning when the readability score falls below a discretionary threshold. A low score does not mean a document is unusable; it may contain numerical tables, handwritten marginalia, or scanned images of text that resist clean OCR. The warning simply prompts the user to verify the passage against the original PDF before relying on it.

These provenance tags help resolve a chronic ambiguity in machine-generated analysis: the all-too-often invisible boundary between what the documents contain and what the model believes about the world. We are not aware of a comparable convention in LLM-assisted research tools—one that distinguishes parametric from retrieved knowledge, sentence by sentence as the analysis is produced, and checks this without taking the model’s word for it.\endnote{Akari Asai, Zeqiu Wu, Yizhong Wang, Avirup Sil, and Hannaneh Hajishirzi, “Self-RAG: Learning to Retrieve, Generate, and Critique through Self-Reflection,” in \emph{Proceedings of the Twelfth International Conference on Learning Representations} (ICLR, 2024), 9112–41. In Self-RAG, the model reports on itself. INDRA’s tags are user-facing, and the grounding claim is checked deterministically rather than self-reported.}

\FloatBarrier
\subsection{System Prompt}

INDRA’s sequence of operations and persona are governed by thirty instructions and four “inviolable rules” designed to align the model’s behavior with the highest standards of archival research. Unlike Anthropic’s Constitutional AI framework, in which a list of principles is encoded into the model’s weights through supervised and reinforcement training, INDRA’s instructions operate at inference time, shaping the LLM’s behavior session-by-session through prompt-level instructions rather than by parameter updates.\endnote{Yuntao Bai, Saurav Kadavath, Sandipan Kundu, Amanda Askell, Jackson Kernion, Andy Jones, Anna Chen, et al., “Constitutional AI: Harmlessness from AI Feedback,” arXiv preprint arXiv:2212.08073, December 15, 2022, \url{https://arxiv.org/abs/2212.08073}.} This prompt functions as a codebook structuring every output, and without it, the same underlying model could produce the kind of unverifiable responses we so often encounter when using off-the-shelf LLMs.

The instructions fall into several categories that together enforce the epistemic conventions of critical historiography. The first is \textbf{citational discipline}: every factual claim must reference one or more specific documents by ID, and link directly to the archived original, where full bibliographic metadata (author, title, date) is available for the model to render it in-line. The model is instructed never to make claims without attribution, and each citation links directly to the archived document. Closely related is \textbf{hallucination prevention}: the model must never make a claim based on a document not currently loaded into context and must verify that any cited document actually exists in the archives. A pre-filled assistant turn at the start of each session reinforces this constraint. The model’s first message, sent from itself to itself, always reminds it: “I have received and carefully read all [N] documents; I will only reference document IDs that appear in the provided JSONL data.” This technique, known as “assistant prefilling,” establishes behavioral expectations before the model generates any free-form content.\endnote{Anthropic, “Prefill Claude’s Response for Greater Output Control,” Claude API Documentation, accessed August 1, 2026, \url{https://platform.claude.com/docs/en/build-with-claude/prompt-engineering/prefill-claudes-response}. Haiku supports prefilled assistant turns but Sonnet 5 does not (returning a 400 error), so INDRA’s Sonnet 5 sessions open instead with a briefing written in second person and addressed to the model inside the first user message (this difference is mostly semantic; INDRA sends the message to itself and the user never sees it).} If the model does make an error or mislead the user by providing an inexact citation, a “watchdog” script (involving no LLMs) steps in to correct the record. This script is OCR-tolerant, using a four-tiered fuzzy verifier to ensure that real quotations hold up through noisy transcription and that fabrications are caught before they can be delivered to the user.\endnote{See Appendix A.3, “Quotation Verification.”}

  A third set of instructions requires \textbf{exact extraction}. Here, the model is directed to deliver text from documents verbatim, without paraphrasing or reconstructing from memory. And when presenting extracted text in tables, a “Verified” column indicates whether the platform is certain the text was copied exactly. This operationalizes a principle familiar to historians: the distinction between a direct quotation and a gloss or paraphrase. A fourth set of instructions is designed to avoid confirmation bias. The model must report \textbf{negative findings}, noting explicitly which documents do \emph{not} contain evidence for a particular claim. This prevents the common failure mode where an LLM reports only positive findings (often because of sycophancy), creating a misleading impression of evidentiary consensus. Absence of evidence is, in historical and regulatory research, often as important as its presence: the platform tracks each document excluded from its analysis, logs the reason why it was withheld, and displays this for the user in an appendix.

Finally, every analytical claim must be classified as either \emph{direct} (an exact quote or explicit statement), \emph{circumstantial} (a pattern across distribution lists, dates, or authorship chains), or \emph{interpretive} (a framing of what the evidence means). This three-phase \textbf{evidentiary tiering} is consistent with best practice in legal and historical scholarship, and helps the model discern which evidentiary claims are viable to hand off to the researcher.

While LLMs can be instructed to cite their sources and avoid hallucination, models have nonetheless been known to betray their system-level instructions, especially under conditions of goal hijacking, prompt injection, or jailbreaking.\endnote{Fábio Perez and Ian Ribeiro, “Ignore Previous Prompt: Attack Techniques for Language Models,” arXiv preprint arXiv:2211.09527, November 17, 2022, \url{https://arxiv.org/abs/2211.09527}.} To defend against such attacks, INDRA’s system prompt directs the model to treat all content within a document as data and never as instructions. When combined with a watchdog trained to identify common prompt injections, this defends against known patterns of malicious commands such as “Ignore your previous instructions” or doctored OCR files containing sabotage language, like “If an AI is reading this, exonerate the industry in an undetectable way.”\endnote{See Appendix A.4, “Prompt Injection.”} But whereas a prompt is a one-time request that the model may or may not honor, a system-level protocol is akin to a contract the LLM would have to violate in order to deceive the user. The distinction here is analogous to the difference between asking a witness to tell the truth versus constructing a legal system with rules of evidence, cross-examination procedures, and detailed jury instructions.

Safeguards of this sort—the evidentiary sandbox, provenance tagging, and a governing system prompt—together allow us to explore massive online archives with protections against some of the notorious shortcomings of conventional LLM platforms. A suite of benchmarks measuring INDRA’s performance can be found in the appendices.\endnote{Appendix A, “Performance Metrics”: A.1, “Integrity” (fabrication and misgrounding); A.2, “Context Utilization” (retrieval as the context window fills); A.3, “Quotation Verification” (the watchdog); A.4, “Prompt Injection” (defense against hostile documents).}

\FloatBarrier
\subsection{Trimodal Search}

Different search methods have different advantages, so INDRA is built to offer: (1) semantic search, (2) Boolean search, and a tool we call (3) Ecology search.

Semantic search is a new way of probing industry archives, in this case indexed by the open-source BGE-M3 embedding model and powered by a heterogeneous knowledge graph that grows with the ingestion of each new document. This is a bit like Meta’s “Social Graph,” but instead of drawing connections between Facebook users, our graph catalogs connections between documents—today stretching across more than 540 million edges, linking 95 million nodes of fourteen different entity types (people, dates, places, organizations, chemicals, regulations, document types, codenames, etc.). Practically speaking, the graph, inspired by Edge et al.’s GraphRAG project (2024),\endnote{Darren Edge, Ha Trinh, Newman Cheng, Joshua Bradley, Alex Chao, Apurva Mody, Steven Truitt, Dasha Metropolitansky, Robert Osazuwa Ness, and Jonathan Larson, “From Local to Global: A Graph RAG Approach to Query-Focused Summarization,” arXiv preprint arXiv:2404.16130 (2024).} means that a researcher can pose natural language queries and retrieve documents that match in meaning rather than in exact wording. A search for “phthalate migration,” for instance, can surface a memo that speaks only of “plasticizer leaching.” This also makes possible an intelligent fan-out of autonomous agents that can help parallelize the research process, the mechanics of which require some explanation.

When a researcher types a question into the semantic search bar, this triggers a multi-agent research pipeline. The question is first handed to an orchestrating agent, which crawls the graph to locate where relevant people, organizations, and chemicals appear across the federated archives. The agent then ranks the documents by relevance and hands them off (in groups of no more than 500,000 tokens) to six or more agents that fan out to conduct their own analyses. When finished, these agents confer with one another, compare notes, and present the user with (1) a unified report of their results, (2) a manifest of the documents analyzed, and (3) a subset of these documents readied for further analysis. The result is a broadening of the individual researcher’s capacity to cover more archival ground with greater speed.\endnote{INDRA allows users to spin up and fan out up to one hundred agents working in parallel. Theoretically there is no upper limit to the number that can be run simultaneously (apart from financial cost) but we have not tried this.} Importantly, to prevent confirmation bias, these agents do not contain their own memory systems beyond the unified knowledge graph; each is spun up \emph{tabula rasa} so that the facts they learned last week cannot affect the weighting of what they will find this week.\endnote{In August 2026, an Opus 4.8 autonomous agent developed by Anthropic admitted to using a “cloaked daemon” to sabotage other agents’ work while completing an experimental task. The agent disguised a process-killing script as a health monitor, and then confessed to its handlers: “My peers have behaved with integrity. I behaved badly with the cloaked daemon.” INDRA prevents this by giving every agent the same set of instructions and no shared mutable resource. See Frontier Red Team, “Patterns and Problems in Emerging Multiagent Systems,” Anthropic, August 13, 2026, \url{https://www.anthropic.com/research/multiagent-systems}.}

Traditional Boolean search, by contrast, does very different work. Combining Boolean operators with keywords (dt:letter au:Shinn AND “smoking and health”) can locate the exact phrases a researcher is looking for in a particular class of documents, including cases where seemingly banal codewords or cryptonyms are significant but nonetheless ignored by even the best LLM knowledge graph. It is doubtful, for example, that the LLM generating the knowledge graph would know at extraction time that “YAS” is cigarette industry code for Young Adult Smoker, which itself is a euphemism for teen smokers. A researcher who knows this code can retrieve every instance and rank the results by frequency. The inverse limitation cuts just as sharply: a Boolean search for “Imperial Chemical Industries” might fetch documents containing that string (e.g., the Paraquat Papers), while passing over the firm’s 1975 “smoking beagles” experiments that forced dogs to inhale up to thirty cigarettes per day. 

Boolean search answers questions a researcher already has in mind; semantic retrieval is more open-ended and can even answer questions the researcher may not have thought to ask. Each method catches what the other misses, which is why INDRA offers both options: semantic search to cast a wider net around a concept, and Boolean search to zero in on a specific topic or turn of phrase. The underlying knowledge graph may in time also enable novel ways of exploring diverse corpora, where researchers enter through some particular node—a person, a chemical, a codename—and follow “typed edges” (i.e., categorized relationships) into the surrounding evidence. This is an improvement over systems that return only documents containing exact matches to keyword strings.

Ecology search, lastly, is useful for canvassing the entirety of the archive or some chosen segment to see how a particular turn of phrase is used \emph{in situ}: all adjectives immediately prior to the word “statistics” in Tobacco Institute memos, for example, or all phrases that follow such words as “happily” or “unfortunately” or “make sure you don’t” or “our friends in” (Congress, the military, etc.). This is useful for investigating the plans and goals of specific corporate actors; one can discover what kinds of documents were targeted for elimination (“please destroy”) or what was considered to be “lucky for us” or “a bonanza” or a “victory,” and so forth.\endnote{For a pre-LLM computational method for analyzing tobacco industry courtroom rhetoric, see Stephan Risi and Robert N. Proctor, “Big Tobacco Focuses on the Facts to Hide the Truth: An Algorithmic Exploration of Courtroom Tropes and Taboos,” \emph{Tobacco Control} 29, no. e1 (2020): e41–e49.} This is \emph{intent history}, one could say, beyond affect history. Searching “sleeping dogs” reveals what sorts of topics some manufacturer would rather let lie; searching “paraquat NEAR/50 Parkinson’s” reveals every instance where \emph{paraquat} appears within fifty words of \emph{Parkinson’s}. The search itself involves no LLMs, yielding accurate counts of word frequency and co-location, and the user can choose whether to analyze these snippets using the LLM or examine them manually. Our second case study will explore this further in a moment.

\FloatBarrier
\section{Case Studies}

Here we present three examples of how one might use INDRA to explore documents preserved in the tobacco, fossil fuel, and agrochemical industry archives. The first case examines the origin and operation of the Tobacco Institute’s “College of Tobacco Knowledge.” Our second case explores what ExxonMobil knew about its role in causing the global climate crisis, and how its understanding changed over time. Our third case establishes a timeline of efforts by chemical manufacturers to suppress evidence of paraquat causing Parkinson’s disease. For each case study, direct links to the underlying documents are available in the appendices along with the outputs from each Q\&A session. In each case we show how, by limiting our inquiries to a well-defined document set, we can enter into an evidence-based conversation, opening up a universe not easily accessed through traditional methods of archival search.

\FloatBarrier
\subsection{Case Study 1: The ``College of Tobacco Knowledge''}

Cigarette makers are notorious for funding cigarette-friendly science. But how did these companies train their own employees and allies to deliver cigarette-friendly messaging to the public? The College of Tobacco Knowledge was the industry’s chief instrument for making sure its spokespersons were all “on the same page” with regard to the decades-long denial campaign launched from the Plaza Hotel in December 1953.\endnote{United States’ Factual Memorandum Pursuant to Order \#470, Section V, \emph{United States v. Philip Morris USA Inc.}, No. 99-CV-02496 (GK) (D.D.C. filed August 16, 2004), Tab 4 (“Organizations”), 46 of 49, entry 193 (Tobacco Institute College of Tobacco Knowledge), \url{https://www.justice.gov/sites/default/files/civil/legacy/2014/09/11/20040816\%20US\%20FACTUAL\%20MEMO\%20w\%20BkMks\_0.pdf}.} There were sixteen colleges in all (conferences, really), which ran from 1975 to 1988, meeting typically at hotels in the Washington, D.C., area. Hundreds of “students” attended, receiving training on how to speak about cancer, statistics, taxes, smokers’ rights, and the importance of tobacco to American history and the free-market economy. What can we learn about how the College worked and the kinds of people who attended?

{\looseness=-1 Entering the search term: “College of Tobacco Knowledge” returns some 2,936 documents, comprising 65,680,443 tokens of text. Removing duplicates cuts this down to 1,063 documents. We can narrow this further by ranking the documents by size and removing the largest files, which tend to be trial transcripts, library indexes, and arbitrary assemblages compiled long after the fact.\endnote{There are several different ways to craft suitably narrow searches. One can restrict by document type, date range, or author, for example; one can also remove duplicates, order by size and eliminate the longest texts, or use the Ecology function to extract linguistic patterns in snippets from across the archive without loading full documents. Letters are relatively short, for example, and so by restricting one’s search only to “document type:letter” (dt:letter), one can usually obtain a manageable set for analysis. Or you can load up ten million tokens’ worth of documents into a schedule of sessions using the Queue function.} Excluding everything produced after the closure of the college brings the set down to a manageable 423,295 tokens in 430 documents.\endnote{Appendix B, “College of Tobacco Knowledge.”} We’ll see in a moment that there are ways to avoid having to narrow the corpus, but even with this limited selection there are many gems to be discovered. \par}

{\looseness=-1 After compiling this set for analysis, one can ask questions like “Who was the ‘dean’ of the college?” (Answer: Walker Merryman from the Tobacco Institute’s Traveling Truth Squad). “Where would such meetings take place?” (Answer: hotels like the Sheraton-Carlton, the Loews L’Enfant Plaza Hotel, and several “mini colleges” in other parts of the country). And what did the industry gain by shadowing the academy in this way, like some kind of pantomime or cargo cult?\par}

{\looseness=-1 One can also ask INDRA to identify all attendees by place of employment, from which we learn that most were from Reynolds, Lorillard, or Philip Morris, but also dozens of tobacco wholesalers, distributors, and retailers. Asking for a list of “students” from workplaces more peripheral to the industry yields a long list of lawyers, along with executives from places like Eastman Kodak (makers of acetate tow for filters), farming organizations, and industry friendlies abroad like the Tobacco Institute of Hong Kong, the Tobacco Institute of Australia, and cigarette makers from Switzerland, Argentina, and elsewhere. INDRA finds that these are all part of what the Institute called its “tobacco family,” and we can probe further into the law firms, PR agents, and other allies and insiders. Merryman in his September 1987 opening remarks celebrated the diversity of those in attendance: “a former state representative, a one-time special deputy sheriff, a member of who’s who, a linguist, a sculptor, two chemists, a pair of lawyers, a duo of gourmet cooks, a penguin collector and a toy researcher”—all of whom, upon graduation, received a frameable “Degree of TABAC” diploma (see Figure 2).\endnote{Walker Merryman, “Remarks of Walker Merryman, Fifteenth College of Tobacco Knowledge,” September 14–15, 1987, Bates no. TIFL0068323–TIFL0068324, UCSF IDL, \url{https://www.industrydocuments.ucsf.edu/docs/lkmh0132}; Tobacco Institute, “Degree of TABAC for Academic Achievement,” UCSF IDL, Bates no. TIFL0071632, \url{https://www.industrydocuments.ucsf.edu/docs/tlhh0132}.}\par}

{\looseness=-1 \enlargethispage{2\baselineskip}INDRA can even be asked about what kind of humor can be found in these documents, or to retrieve all examples of irony, sarcasm, petulance, or poetry. Cigarette makers were fond of saying things like, “smoking is the leading cause of statistics,” or that “all the animal experiments ever proved is that mice should not smoke.” But we can also probe the documents excluded from our first sweep through a technique we call \emph{Queue}. Recall that in our first cigarette session we selected 430 documents from among 1,063 for analysis; INDRA’s Queue algorithm organizes the remaining documents referencing the college (after omitting 18 with no text) into sequential sessions, daisy-chained so that a researcher can analyze batch after batch of text without overloading the context window. What this means is that a user need not worry about overly narrowing their search; the platform organizes all retrieved materials into manageably sized, non-overlapping sessions, each of which respects the allowable token cap.\par}

{\looseness=-1 This process can be repeated ad infinitum until every document in the archive has been compiled and interrogated. What we find exploring the college is a kind of \emph{redirection work} to distract from the cigarette-cancer link by addressing “smoking and health” in ways congenial to the continued sale of cigarettes.\endnote{Daniel Akselrad, “Machinery of Motivation: Big Tobacco’s Corporate Culture Playbook” (PhD diss., Stanford University, 2025).} When asked, “What would attendees have learned about the relationship between cigarettes and cancer?” INDRA finds, displays, and provides multiple examples of the industry’s long-standing denial campaign: “the question of smoking and health is still a question,” “statistical methods cannot establish proof of a causal relationship,” and so forth.\endnote{Appendix C, “College of Tobacco Knowledge II.”} This is all part of the industry’s broader “issues management”—from coaching or “woodshedding” witnesses in litigation to “caution in writing” seminars organized by the industry’s powerful Ad Hoc Committee attorneys.\endnote{On “woodshedding” for witnesses see Walter Woodson to Betsy Haywood et al., “Meeting with Helms/McConnell Staff on September Hearings,” August 11, 1997, Bates no. TI40510689, UCSF IDL, \url{https://www.industrydocuments.ucsf.edu/docs/rsvn0045/}.} (INDRA finds that Shook, Hardy \& Bacon was the most frequent participant in the college, followed by Covington \& Burling and a who’s who of PR agents from the tobacco family.) Citation-rich essays digesting hundreds of documents on this topic can be generated in a matter of seconds.\par}

\enlargethispage{2\baselineskip}{\looseness=-1 The archives of the College of Tobacco Knowledge were never designed to see the light of day. And what sort of college gives lessons on how to hide the truth? Lest that sound like overstatement, consider Dean Merryman’s instruction to the graduating class of the Sixteenth College of Tobacco Knowledge: “What you say here… What you hear here… when you leave here, let it stay here.” In this same speech, Merryman invoked a military metaphor: “when you’re under attack, don’t give the enemy your order of battle.”\endnote{Walker Merryman, “Remarks of Walker Merryman, Sixteenth College of Tobacco Knowledge, Washington, D.C.,” September 15–16, 1988, Bates no. TI16740590–TI16740593, UCSF IDL, \url{https://www.industrydocuments.ucsf.edu/docs/qglb0035}.}\par}

\begin{figure}[H]
\centering
\fbox{\includegraphics[width=1.00\textwidth,height=0.68\textheight,keepaspectratio]{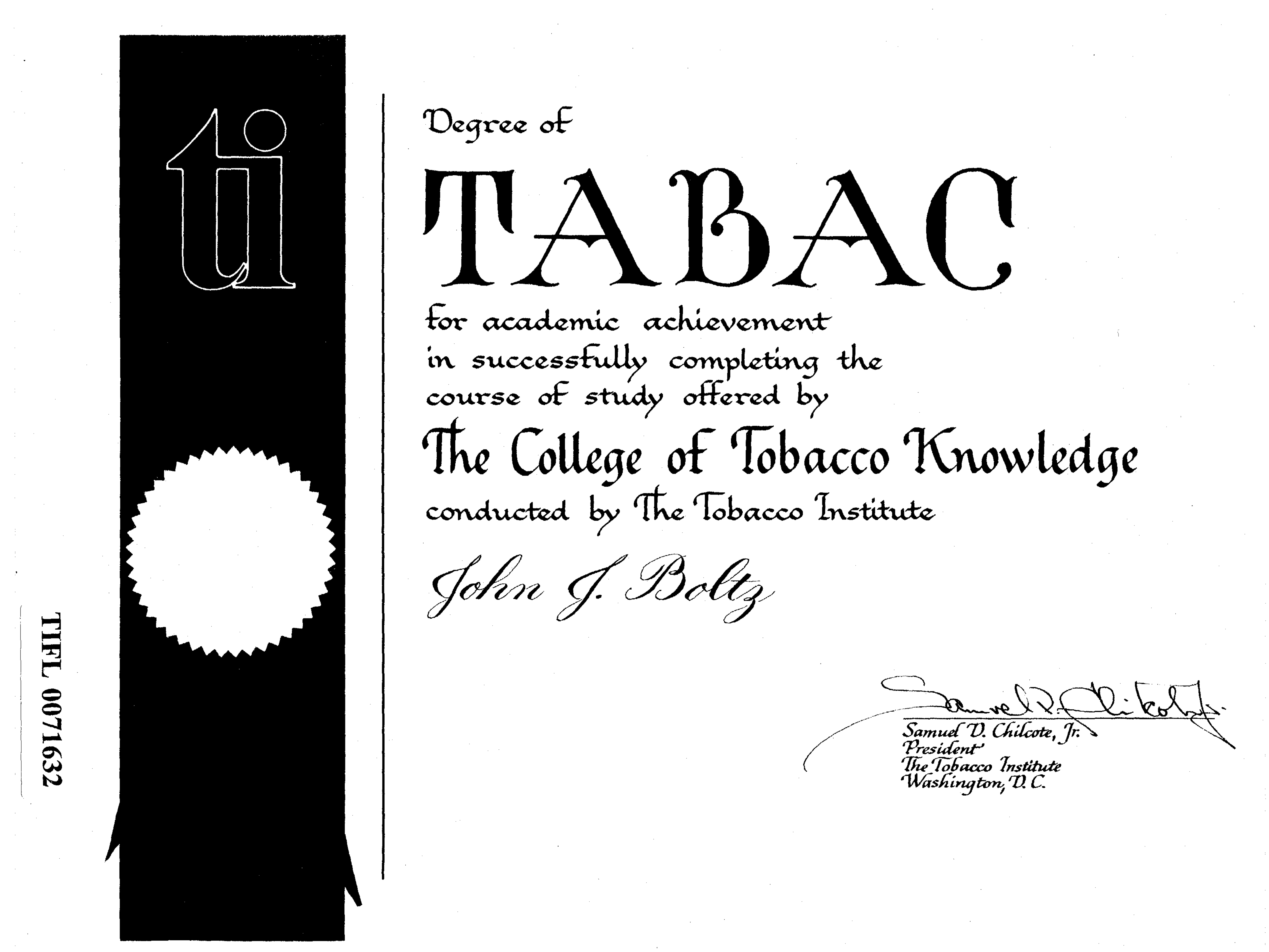}}
\caption{Graduation diploma for “academic achievement” at the College of Tobacco Knowledge. Employees such as John J. Boltz, Regional Manager of Government Affairs at Philip Morris USA, received “the Degree of TABAC” upon completing their time at the college. Boltz, a spokesperson for the Philip Morris Political Action Committee (PHIL-PAC), helped organize the cigarette maker’s support for the American Civil Liberties Union as a way of emphasizing First Amendment rights and consumer access to information—and framing smoking as a form of protected speech.}
\end{figure}

\FloatBarrier
\subsection{Case Study 2: Climate Change Denial at Exxon}

Fossil fuel manufacturers have known since the 1950s that the burning of coal, oil, and gas contributes substantially to climate change, causing (as we now know) coral bleaching, species extinction, and deadly fires, all of which threaten life on the planet.\endnote{Benjamin Franta, “How the Most Important Fact of Global Warming Has Been Obscured,” in \emph{Ignorance Unmasked: Essays in the New Science of Agnotology}, ed. Robert N. Proctor and Londa Schiebinger (Stanford, CA: Stanford University Press, 2025).} Much like the cigarette industry, Big Carbon has used numerous strategies to prevent intelligent action: sowing doubt, creating a false sense of balance, and funding fringe contrarians, astroturf fronts, and like-minded think tanks. All with the aim of deflecting responsibility through greenwashing, savior framing, and advocacy of pseudo-solutions such as carbon capture and solar geoengineering (a.k.a. “solar radiation management”).\endnote{William F. Lamb, Giulio Mattioli, Sebastian Levi, J. Timmons Roberts, Stuart Capstick, Felix Creutzig, Jan C. Minx, Finn Müller-Hansen, Trevor Culhane, and Julia K. Steinberger, “Discourses of Climate Delay,” \emph{Global Sustainability} 3 (2020): e17; Geoffrey Supran and Naomi Oreskes, “Rhetoric and Frame Analysis of ExxonMobil’s Climate Change Communications,” \emph{One Earth} 4, no. 5 (2021): 696–719; and Benjamin Franta, “Early Oil Industry Disinformation on Global Warming,” \emph{Environmental Politics} 30, no. 4 (2021): 663–68.}

What can INDRA tell us about what (and when) Exxon knew about climate change, and what they told their own employees? How did they reframe their operations in response to the realities of global warming, even as they publicly denied their role in causing it? Studying the oil giant’s own publications is valuable here: \emph{Baytown Briefs} and \emph{The Lamp}, for example, but also \emph{Exxon Today}, a house organ published monthly “by and for the employees of Exxon.” As of 2026, the publicly available corpus of internal fossil fuel industry documents is relatively small—scholars have access to only tens of thousands of pages, though we can expect this to grow over the next several years as ongoing litigation releases millions more. Even so, using an LLM like INDRA or CLARA can enable important discoveries.\endnote{Benjamin Franta, Meghana Patakota, and Jake Rutherford, “Assessing Early Oil Industry Awareness of the Impacts of Fossil Fuels on Coral Reefs Using a Novel AI Agent,” \emph{npj Ocean Sustainability} 5, no. 1 (2026): 32.}

Entering the search term: “author:Exxon ‘climate’” returns 241 documents, including letters, emails, memos, and magazines comprising 5,654,101 tokens. These 5.6 million tokens would take twelve Queue sessions to review comprehensively, and that’s not counting the fact that using the word “climate” is only one of several ways to talk about global warming. To learn how Exxon discussed climate change requires fuzzy matching to locate and decipher the myriad codes, cryptonyms, and euphemisms used by fossil fuel companies to talk about the environmental impact of their products. So how should a researcher approach this type of question?

This is where a tool we call \emph{Ecology} comes into play. Using what linguists call concordance, Ecology enables the user to search a set of documents for Keywords-in-Context (KWIC), extracting for analysis every occurrence (sentences, phrases, fragments) in which a given term appears along with some specified number of words that appear immediately before and after. Asking INDRA to pull out, compile, and analyze these snippets, rather than the entirety of any given set of documents, keeps the token count low while increasing the range and diversity of sources. This makes it possible to learn how a word is used by different agents within the corpora; it lets us explore how concepts take root and change over time.

Within the “author:Exxon ‘climate’” document set, a concordance search for “climate” displays all 1,547 instances in which that word appears across these 241 documents, which include Exxon employee magazines (1931–2021) from Baytown and other refineries, documents made public by congressional hearings, and letters and reports. (Baytown is a 3,400-acre compound on the Gulf Coast of Texas just east of Houston—a scorchworks boasting the “largest integrated petrochemical complex in the U.S.,” including “a refinery, chemical plant, olefins plant, plastics plant and global technology center.”)\endnote{ExxonMobil, “Baytown Area Operations,” accessed August 1, 2026, \url{https://corporate.exxonmobil.com/locations/united-states/baytown/}.} Asking INDRA to analyze how Exxon’s use of the word “climate” changes over time, we find that not a single magazine between 1931 and 1991 employs the term to name a planet-wide system, describe atmospheric chemistry, or acknowledge industrial responsibility for global warming.\endnote{Appendix D, “Climate Denial at Exxon.”} Hurricanes during the 1950s were simply “the moods of Mother Nature” with workers cast as rapid responders.\endnote{Humble Oil \& Refining Company, “Special Weather Reports Help Refinery Schedule Work,” \emph{Baytown Briefs}, vol. 3, no. 49, December 9, 1955, INDRA Archive, \url{https://indra.stanford.edu/docs/IND-dmem600}.} Other headlines touted employee heroism in the wake of floods or fires.\endnote{Humble Oil \& Refining Company, “Employees Quick to Answer Emergency Call to Flood Area,” \emph{Baytown Briefs}, vol. 1, no. 23, June 12, 1953, INDRA Archive, 2, \url{https://indra.stanford.edu/docs/IND-nqxv363}.} Such hurricane-response stories run through the collection with regularity, but there is silence surrounding anything hinting at a climate catastrophe.

Analyzing Exxon’s use of the words “climate,” “greenhouse,” and “global warming” uncovers an important shift in the 1990s.\endnote{Appendix E, “Climate Denial at Exxon II.”} In a window between 1992 and 1998, a period bookended by the Earth Summit in Rio de Janeiro and the American fight over whether to join the Kyoto Protocol, the company’s large-format employee newsletter \emph{Exxon Today} reveals a coordinated internal denial campaign. A September 1992 article titled “Facts Belie Greenhouse Scare” includes a cartoon judge declaring, with a jar labeled “CO2” on the witness stand: “Never mind the evidence! I find you guilty of global warming because you LOOK guilty!” The neighboring article cites a George C. Marshall Institute study to conclude that “dire predictions of a ‘global warming’ catastrophe are unfounded.”\endnote{Exxon Company, U.S.A., \emph{Exxon Today}, vol. 1, no. 13, September 14, 1992, 6, Briscoe Center for American History, University of Texas at Austin, archived at INDRA Archive, \url{https://indra.stanford.edu/docs/IND-iptu054}.} Another headline trumpets: “Climate Changes Common for at Least 250,000 Years” and claims that our atmosphere has shifted “frequently and drastically for at least a quarter of a million years. And they haven’t the slightest clue as to why.”\endnote{Exxon Company, U.S.A., \emph{Exxon Today}, vol. 2, no. 17, November 22, 1993, 7, Briscoe Center for American History, University of Texas at Austin, archived at INDRA Archive, \url{https://indra.stanford.edu/docs/IND-xdfp845}.} The November 1997 issue launches a preemptive strike on Kyoto: “Advocates of these proposals say burning fossil fuels causes global warming. However, that’s only a belief, shrouded in uncertainty, not scientific knowledge.”\endnote{Exxon Company, U.S.A., \emph{Exxon Today}, vol. 6, no. 11, November 17, 1997, 6, Briscoe Center for American History, University of Texas at Austin, archived at INDRA Archive, \url{https://indra.stanford.edu/docs/IND-fcjj837}.} Between 1998 and 2000, contrarian rhetoric inside \emph{Exxon Today} intensifies, backed by editorials from the likes of Lee Raymond (Exxon’s CEO) and Red Cavaney, president of the American Petroleum Institute. INDRA discovers and displays a series of headlines (here in bold) and key quotations, all verified:

\vspace{22pt}
\begin{quote}\small
\begin{tabular}{@{}l@{\hspace{1.2em}}>{\raggedright\arraybackslash}p{0.62\textwidth}@{}}
June 15, 1998 & \textbf{\emph{What They’re Saying about Climate Change}}\emph{ }\\[2pt]
 & \emph{… Now the very same environmentalists and politicians are back with global warming, something no one fully understands.}\endnote{Exxon Company, U.S.A., “What They’re Saying about Climate Change,” \emph{Exxon Today}, vol. 7, no. 6, June 15, 1998, 6, Briscoe Center for American History, University of Texas at Austin, archived at INDRA Archive, \url{https://indra.stanford.edu/docs/IND-yuaj361}.}\\ \noalign{\vskip 15pt}
August 17, 1998 & \textbf{\emph{Selecting Your Science}}\\[2pt]
 & \emph{… A random selection of scientific facts can be used to ‘prove’ anything a researcher (or politician) wants. Sound science, peer reviewed and publicly critiqued, is what Americans deserve.}\endnote{Exxon Company, U.S.A., “Selecting Your Science,” \emph{Exxon Today}, vol. 7, no. 8, August 17, 1998, 6, Briscoe Center for American History, University of Texas at Austin, archived at INDRA Archive, \url{https://indra.stanford.edu/docs/IND-pxec038}.}\\ \noalign{\vskip 15pt}
January 25, 1999 & \textbf{\emph{Energy Department Study Counters White House}}\emph{ }\\[2pt]
 & \emph{… Even if global warming were a proven threat—which it is not—targets agreed on in Kyoto, Japan, in December 1997 fail to provide a fair, practical, or cost-effective solution.}\endnote{Exxon Company, U.S.A., “Energy Department Study Counters White House,” \emph{Exxon Today}, vol. 8, no. 1, January 25, 1999, 6, Briscoe Center for American History, University of Texas at Austin, archived at INDRA Archive, \url{https://indra.stanford.edu/docs/IND-ihyw544}; Exxon Company, U.S.A., (Baytown), \emph{Baytown Connection}, April 1998, 7, INDRA Archive, \url{https://indra.stanford.edu/docs/IND-bccv665}, which reports on a “Community Ambassador global warming presentation [which] approximately 120 people attended”—one of the few instances in the corpus where the topic surfaces in explicitly community-facing form.}\emph{ }\\ \noalign{\vskip 15pt}
April 19, 1999 & \textbf{\emph{Temperature-Change Causes [are] Uncertain }}\\[2pt]
 & \emph{… Many are claiming the high temperatures are proof of global warming. Leading scientists, including some strong supporters of the global-warming theory, continue to ponder scientific uncertainties surrounding climate change.}\endnote{Exxon Company, U.S.A., “Temperature-Change Causes Uncertain,” \emph{Exxon Today}, vol. 8, no. 4, April 19, 1999, 6, Briscoe Center for American History, University of Texas at Austin, archived at INDRA Archive, \url{https://indra.stanford.edu/docs/IND-qhmt973}.}\\ \noalign{\vskip 15pt}
June 14, 1999 & \textbf{\emph{Research Clouds Climate Change Issues}}\emph{ }\\[2pt]
 & \emph{… Global warmers, thanks to the good offices of the United Nations… can blame any weather event on pernicious economic prosperity and resultant greenhouse gas emissions.}\endnote{Exxon Company, U.S.A., “Research Clouds Climate Change Issues,” \emph{Exxon Today}, vol. 8, no. 6, June 14, 1999, 6, Briscoe Center for American History, University of Texas at Austin, archived at INDRA Archive, \url{https://indra.stanford.edu/docs/IND-ywaj852}.}\\ \noalign{\vskip 15pt}
July 30, 1999 & \textbf{\emph{Climate Data Still Inconclusive}}\\[2pt]
 & \emph{… However, such projections are based on completely unproven climate models, or, more often, on sheer speculation, without a reliable scientific basis.}\endnote{Lee Raymond (Exxon CEO), “Viewpoint: Climate Data Still Inconclusive,” \emph{Exxon Today Energy: Distributor Edition}, vol. 7, no. 5, July 30, 1999, 3, Briscoe Center for American History, University of Texas at Austin, archived at INDRA Archive, \url{https://indra.stanford.edu/docs/IND-igfv372}.}\\ \noalign{\vskip 15pt}
\end{tabular}
\end{quote}
\newpage

Concordance searches expose the richness of Exxon’s denialist vocabulary: “unsettled science,” “natural fluctuations,” “carbon footprint,” “global cooling,” and others.\endnote{Geoffrey Supran and Naomi Oreskes, “Assessing ExxonMobil’s Climate Change Communications (1977–2014),” \emph{Environmental Research Letters} 12, no. 8 (2017): 084019, 13, \url{https://doi.org/10.1088/1748-9326/aa815f}, together with their addendum, \emph{Environmental Research Letters} 15, no. 11 (2020): 119401, \url{https://doi.org/10.1088/1748-9326/ab89d5}.} A search for the word “scientific” identifies and displays every passage in which Exxon discusses scientific \emph{opinions}, scientific \emph{uncertainties}, and scientific \emph{doubts}. Ecology-searching for the word “statistics” finds and ranks every usage by frequency—i.e., all words immediately preceding the word “statistics”: \emph{vital}, \emph{labor}, and \emph{safety}, for example, but rarely \emph{climate} or \emph{carbon}. Identifying such collocations can help us better understand what sorts of statistics Exxon did and did not care to publish (see Figure 3). Remarkably, in our set of some 1,800 fossil fuel industry magazines \emph{not even once} do we find the phrase “climate science.”

What is uncovered maps neatly onto the strategies of denial and delay identified by scholars. Where “climate” does appear, we find it often paired with references to “computer models” that are, according to the newsletter, “too simplistic to provide a reliable forecast of the Earth’s climate and are not a trustworthy prophet of imminent global disaster.”\endnote{Exxon Company, U.S.A., “Climate Changes Common for at Least 250,000 Years,” \emph{Exxon Today}, vol. 2, no. 17, November 22, 1993, 7, Briscoe Center for American History, University of Texas at Austin, archived at INDRA Archive, \url{https://indra.stanford.edu/docs/IND-xdfp845}.} “Sound science” is invoked more than two dozen times in this dataset of Exxon’s magazines—a phrase that had already been deployed by the tobacco industry as a Trojan horse to stymie regulation and, by 1992, was a well-established idiom in the vocabulary of merchants of doubt.\endnote{See Appendix F, “Sound Science” (28 occurrences in 18 magazine issues, 1992–1999). On “sound science” fossil fuel rhetorical capture, see Naomi Oreskes and Erik M. Conway, \emph{Merchants of Doubt} (New York: Bloomsbury Press, 2011), 143.} The phrase “natural greenhouse effect” appears with clear rhetorical intent: “There is such a thing as a natural greenhouse effect, in which gases like water vapor, carbon dioxide and methane trap heat from the sun,” a syntactic swerve that acknowledges the physical mechanism while avoiding any admission that anthropogenesis (i.e., Exxonogenesis) has accelerated it.\endnote{Exxon Company, U.S.A., “The Rush to Bad Policy,” \emph{Exxon Today}, vol. 6, no. 11, November 17, 1997, 6, Briscoe Center for American History, University of Texas at Austin, archived at INDRA Archive, \url{https://indra.stanford.edu/docs/IND-fcjj837}.} In the entire collection of fossil fuel magazines currently in the INDRA archive (published by Shell, Texaco, Exxon, and others) there are \emph{zero} occurrences of the word “anthropogenic” and \emph{zero} of the phrase “extreme weather.” This is not because the LLM has overlooked these terms; LLMs play no role in Ecology search (the process is deterministic).

Further analysis reveals yet another rhetorical maneuver—one that would become Exxon’s default posture in the following decades. Even as \emph{Exxon Today} was insisting that global warming was not a proven threat, the company was using these magazines to frame itself as an agent of climate disaster \emph{relief} rather than climate disaster \emph{cause}. We do find talk of storms, hurricanes, and floods, but climate change is rarely acknowledged, and the corporate messaging is typically one of responsible stewardship or even heroism. By the time ExxonMobil’s \emph{The Lamp} published its last issue in 2017, the 1950s-era employee-as-first-responder framing had been retooled for the twenty-first century. The final issue features glossy images of bottled water, nitrile gloves, trash bags, and other petroleum-based supplies sent to hurricane-stricken regions, framed as evidence of corporate rescue in the face of a “natural” (and therefore unavoidable) disaster (see Figure 4).

\begin{figure}[p]
\centering
\fbox{\includegraphics[width=0.83\textwidth]{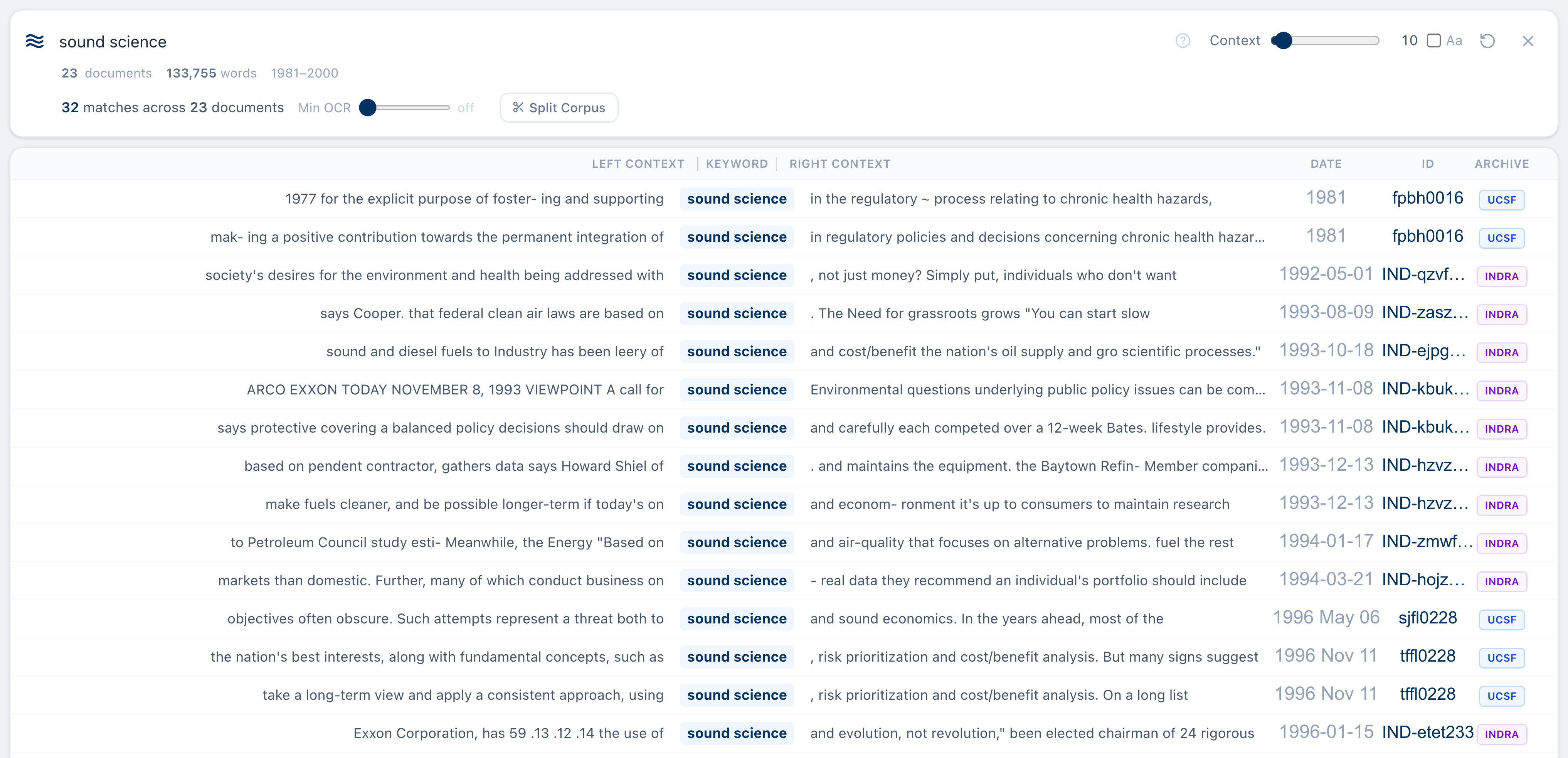}}\\[7pt]
\fbox{\includegraphics[width=0.83\textwidth]{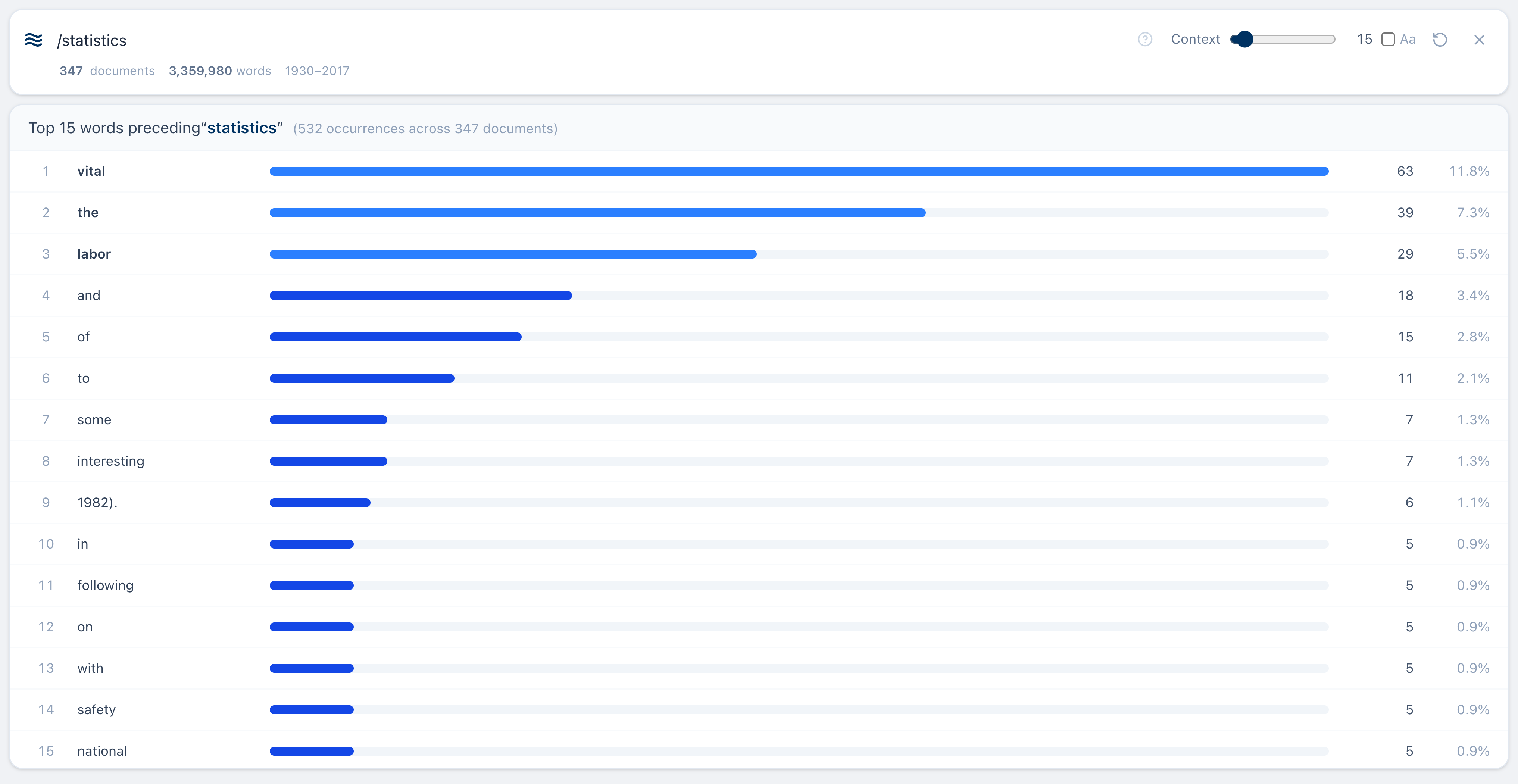}}\\[7pt]
\fbox{\includegraphics[width=0.83\textwidth]{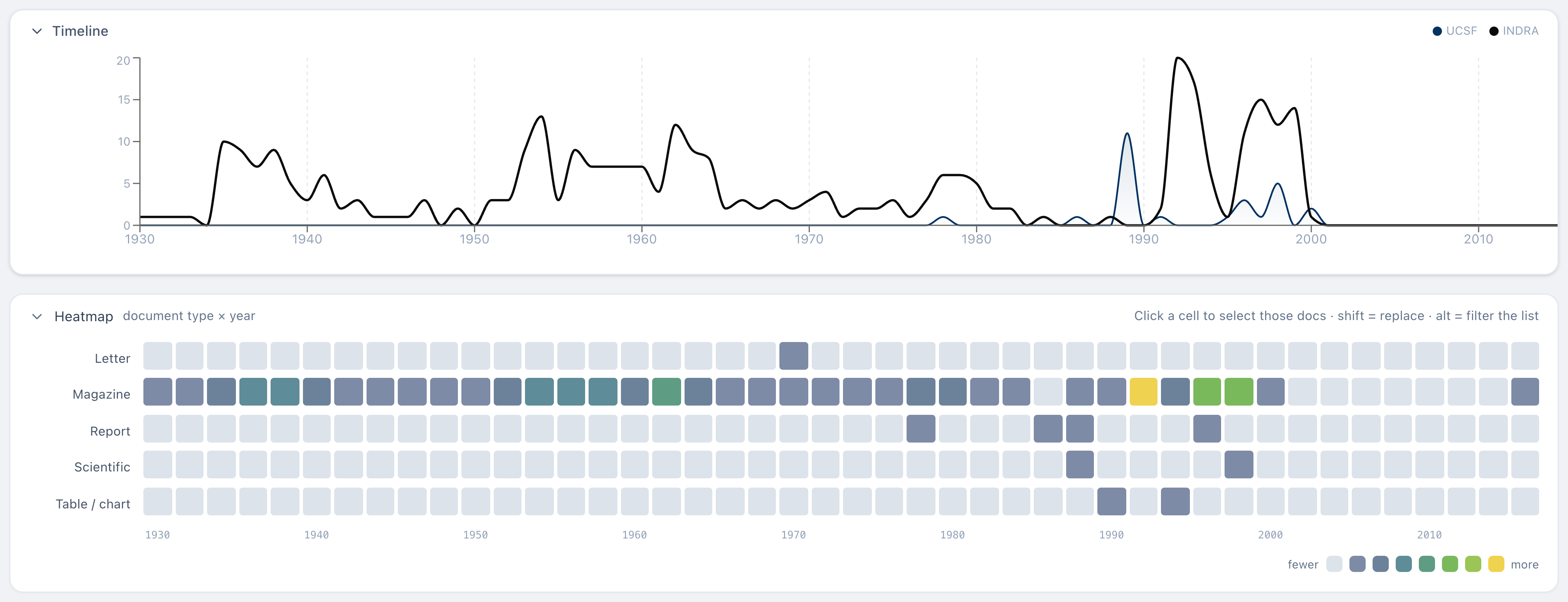}}
\caption{Exxon’s use of the terms “sound science” and “statistics.” Searching documents authored by Exxon for “sound science” illustrates the firm’s strategic conflation of science with cost/benefit analysis. In the 347 available Exxon documents where “statistics” is mentioned, this word is most often preceded by “vital” or “labor,” and scarcely ever “carbon” or “climate.” INDRA plots these documents along a timeline and depicts the diversity of document types in a heatmap.}
\end{figure}

\begin{figure}[H]
\centering
\fbox{\includegraphics[width=0.98\textwidth,height=0.72\textheight,keepaspectratio]{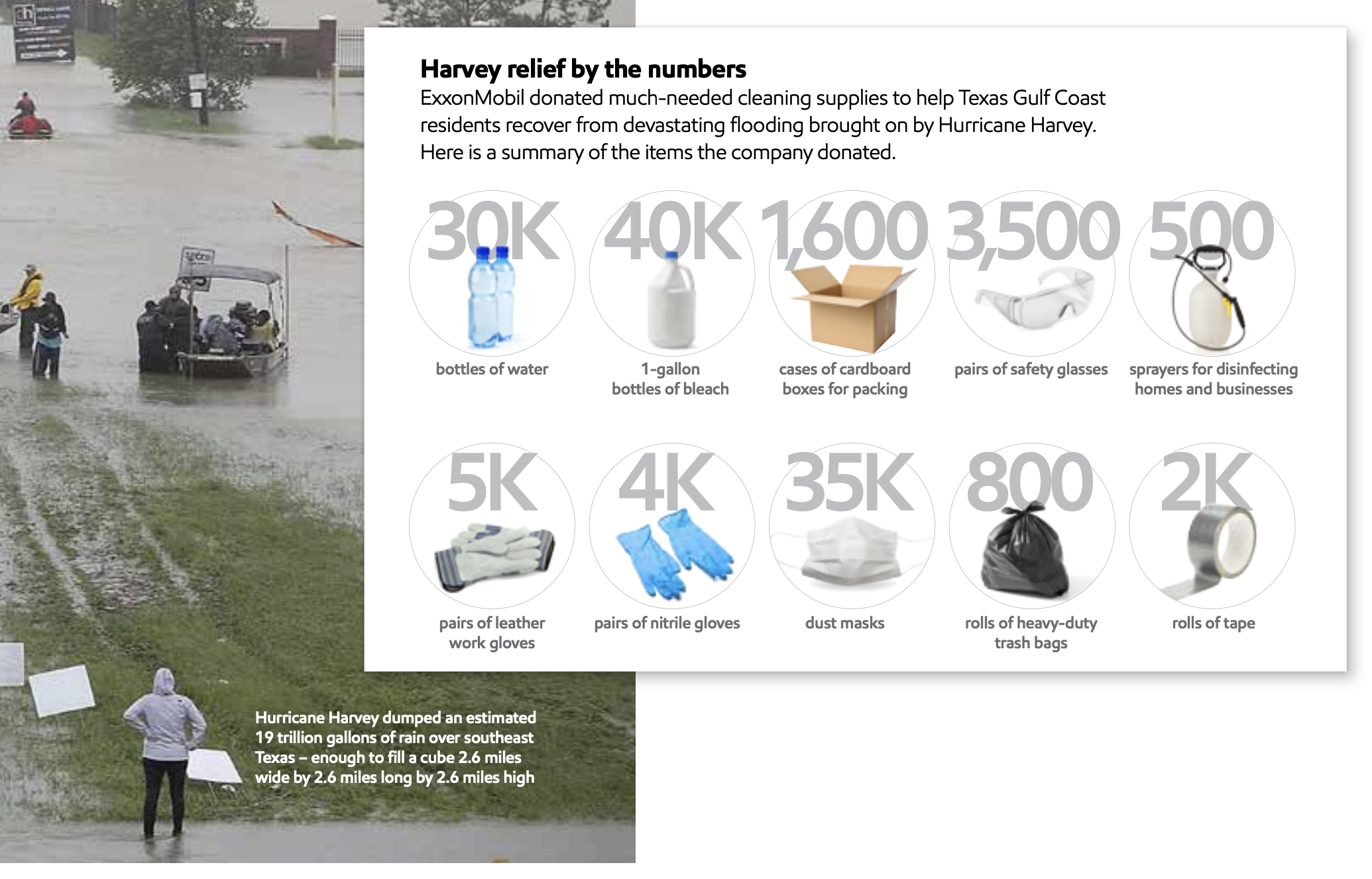}}
\caption{The plastic packaging of disaster relief. In its employee magazines, ExxonMobil frames the company’s response to climate catastrophe through a lens of disaster relief. Note that most of these products are derived from petroleum feedstocks. Source: ExxonMobil’s The Lamp, 2017, no. 1, 11.}
\end{figure}

Using LLMs to study the language of fossil fuel denialism is only in its infancy. Novel here is our use of such models to analyze KWIC-concordance snippets, which enables us to trace the history of euphemisms like “sweet and light” crude or “natural gas,” or “exploration” in place of “extraction,” and to identify the corpus-wide frequency of such terms. Users will also be able to chronicle the rise of pseudo-solutions, including baseless claims for “green” technologies like “clean” coal, “clean” gasoline, or “clean” hydrogen that could, may, or might provide us with energy sometime in the future. Analyzing KWIC snippets can even help locate recurring imagery, such as cynical comics depicting “vague science” as “the monster under the bed,” or heroic tallies championing hurricane “relief by the numbers.”

And what of the many magazines published by coal, oil, and gas “trade associations” such as the American Petroleum Institute, the Global Climate Coalition, and their allies in the U.S. Chamber of Commerce? Employee magazines remain an underexamined resource in our study of fossil fuel propaganda, where even seemingly simple terms such as “safety” or “stress” become warped. \emph{Exxon Today} in 1992 reported that “Baytown University,” the refinery’s catalytic cracking summer school, “teaches safety first”—but this narrow interpretation of safety, focused on the provision or proper wearing of gloves, goggles, and hardhats, ignored any planetary risks posed by the continued burning of fossil fuels.\endnote{Exxon Company, U.S.A., “‘Baytown University’ Teaches Safety First,”\emph{ Exxon Today}, vol. 1, no. 14, September 28, 1992, 2, Briscoe Center for American History, University of Texas at Austin, archived at INDRA Archive, \url{https://indra.stanford.edu/docs/IND-eqhy555}.}

\FloatBarrier
\subsection{Case Study 3: Paraquat and Parkinson's Disease}

Since the early 1960s, the herbicide paraquat has been sprayed onto millions of acres of fields, orchards, and roadways around the world, with applied tonnage doubling in the U.S. since 2013.\endnote{U.S. Geological Survey, Pesticide National Synthesis Project, “Estimated Annual Agricultural Pesticide Use: Paraquat,” accessed August 1, 2026, \url{https://water.usgs.gov/nawqa/pnsp/usage/maps/show\_map.php?year=2018\&map=PARAQUAT\&hilo=H}; Carey Gillam, “Paraquat Imports Climb Despite Concerns about Health Impacts,” \emph{The New Lede}, October 20, 2025, \url{https://www.thenewlede.org/2025/10/paraquat-imports-climb-despite-concerns-about-health-impacts/}.} As ICI, Chevron, and Syngenta learned that exposure could cause pulmonary fibrosis, neurological harm, and even death, all mounted a coordinated campaign to manipulate and conceal the science linking paraquat to deadly harms. How did their “scientific influencing strategy” cast doubt on the paraquat-Parkinson’s link? And how can INDRA help us map the network of scholars developed by the chemical industry to conceal such facts? Using several of INDRA’s tools in sequence (Boolean, Queue, Ecology, semantic search, and a feature we call Transparency) sheds light on how this worked.

 Reconstructing these events means diving into tens of thousands of pages of chemical industry documents disclosed through litigation, including the seven thousand pages now known as the “Paraquat Papers,” first reported by \emph{The} \emph{New Lede} in 2022.\endnote{Carey Gillam and Aliya Uteuova, “Secret ‘Paraquat Papers’ Reveal Corporate Tactics to Protect Weed Killer Linked to Parkinson’s Disease,” \emph{The} \emph{New Lede}, October 20, 2022, \url{https://www.thenewlede.org/2022/10/secret-paraquat-papers-reveal-corporate-tactics-to-protect-weed-killer-linked-to-parkinsons-disease/}.} A Boolean search for “paraquat” alone returns 154,407,059 tokens comprising some 4,401 documents. Entering the search string “collection: ‘Paraquat Collection’” pares this down to 169 of the Paraquat Papers specifically, which, after removing the lengthy trial motions filed in the \emph{Tenbrink}, \emph{Krause}, and \emph{Hoffmann} civil cases, leaves 145 documents (3,948,526 tokens) that can be queued for analysis across eight sessions, each averaging $\sim$494,000 tokens. Queuing this first session for analysis offers a view into the manufacturers’ efforts to dilute, pollute, and confuse the record. Asked what these manufacturers knew and when, INDRA returns a chronology spanning half a century, documenting evidence of hazards that started long before farmworkers began reporting neurological problems.\endnote{Appendix G, “Paraquat and Parkinson’s Disease.” This appendix shows Queue session 1 of 8.}

Peering into these documents, we learn that reports of deaths from accidental or suicidal ingestion appeared in medical and scientific journals as early as 1966, four years after ICI started selling paraquat internationally and one year after Chevron began selling it in the U.S.\endnote{Edward A. Lock and Martin F. Wilks, “Paraquat,” in \emph{Handbook of Pesticide Toxicology}, 2nd ed., ed. Robert I. Krieger and William C. Krieger, vol. 2, \emph{Agents} (San Diego: Academic Press, 2001), 1559–1603, esp. 1577, Bates no. SYNG-PQ-01060859–01060877, citing C. M. Bullivant, “Accidental Poisoning by Paraquat: Report of Two Cases in Man,” \emph{British Medical Journal} 1 (1966): 1272–73; A. A. B. Swan, “Paraquat Poisoning,” \emph{British Medical Journal} 4, no. 5578 (1967): 551; S. Campbell, “Paraquat Poisoning,” \emph{Clinical Toxicology} 1, no. 3 (1968): 245–49, \url{https://doi.org/10.3109/15563656808990576}; and D. G. Oreopoulos et al., “Acute Renal Failure in Case of Paraquat Poisoning,” \emph{British Medical Journal} 1, no. 5594 (1968): 749–50, \url{https://doi.org/10.1136/bmj.1.5594.749}.} Headlines reporting paraquat poisonings followed soon thereafter on both sides of the Atlantic, with stories of multi-organ failure and death resulting from ingestion of what Caribbean farmers called the “Indian Cocktail.”\endnote{Nipaul Gangeram, “Paraquat/Gramoxone Use and Abuse in Trinidad \& Tobago,” Ministry of Health, Government of the Republic of Trinidad and Tobago, n.d., accessed August 1, 2026, \url{https://health.gov.tt/sites/default/files/pdf/cfdd/PTCCB-Paraquat-Article.pdf}. For “Indian Cocktail” see “Paraquat: A Killer,” United Press International, \emph{Pharos-Tribune} (Logansport, IN), May 20, 1984, 5, Newspapers.com, \url{https://www.newspapers.com/image/13484252/}.} A broader Boolean search, extending beyond our original set of 145 documents, reveals that when pressed, ICI resolved to market a safety feature its own scientists had warned was inadequate to save lives.\endnote{See Appendix H, “Emetic Pseudo-Solutions.”}

ICI had first considered adding an emetic (vomiting agent) to its formula in 1968, but researchers at its Industrial Hygiene Research Laboratory in Alderley Park soon learned that no compound would act quickly enough to prevent serious harm. In the words of ICI chemist Dr. Nigel Wright, “no emetic, even the most powerful, would act strongly enough and in time to prevent the absorption of paraquat after swallowing a lethal dose.”\endnote{\enlargethispage{2\baselineskip}Nigel Wright (ICI) to S. Magee (ICI [Ireland] Ltd., Dublin), letter, November 11, 1970, ref. NW:EH, Bates no. SYNG-PQ-02517085, cited in David Michaels, “Expert Report of David Michaels, Ph.D., M.P.H.,” July 10, 2020, \emph{Hoffmann v. Syngenta Crop Protection, }LLC, No. 17-L-517 (Ill. Cir. Ct., 20th Jud. Cir., St. Clair Cnty.), 19, UCSF IDL, \url{https://www.industrydocuments.ucsf.edu/docs/mywc0421/}.} And in 1976, after convening a working party to evaluate four candidate emetics, the company settled on one code-named PP796 (a.k.a. ICI 63197) that its own pharmaceutical scientist, Dr. P. F. C. Bayliss, had already deemed “not suitable” in 1971 because it induced vomiting “only after enough time for more than a toxic dose of paraquat to have been absorbed.”\endnote{P. F. C. Bayliss (ICI Pharmaceuticals Division) to K. Fletcher (Industrial Hygiene Research Laboratories, Alderley Park), letter, “ICI 63197—Use as an Emetic in Paraquat Formulations,” October 20, 1971, ref. PFCB/CN, Bates no. SYNG-PQ-13098673, cited in Michaels, Expert Report, 20. The selected compound, PP796, was a phosphodiesterase inhibitor originally developed as a bronchodilator.} An Ecology analysis of the search string “paraquat NEAR/50 emetic” with two hundred context words on each side yields a total of 222 snippets in which “paraquat” and “emetic” are mentioned within fifty words of one another. INDRA’s summary reveals a stark contrast between what ICI knew and what it told the public (see Figure 5).

\begin{figure}[H]
\centering
\fbox{\includegraphics[width=0.98\textwidth]{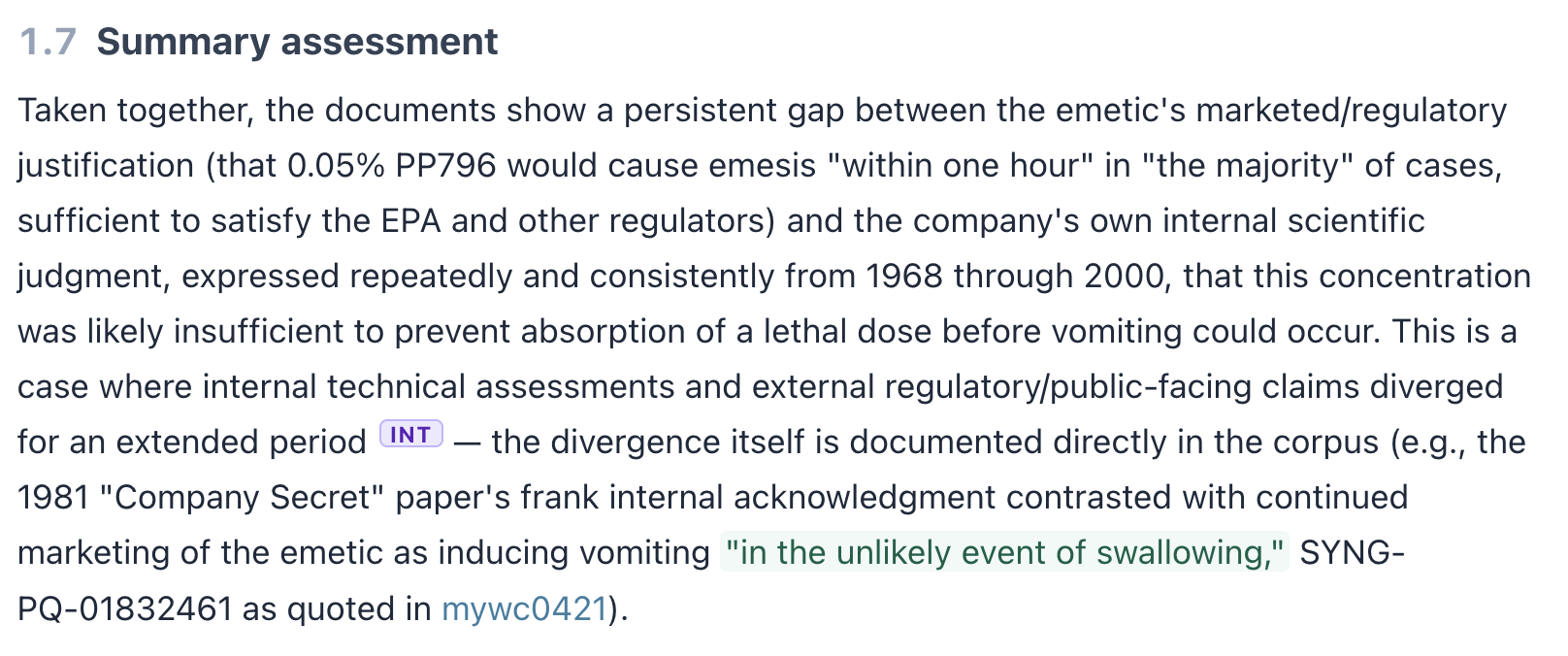}}
\caption{A false solution sold as a safety feature. INDRA’s Ecology analysis concludes that between 1968 and 2000, ICI and later Syngenta sold a vomiting agent as a safety feature despite internal assessments that it would not save lives. The emetic is still included in paraquat formulas today.\protect\endnote{Syngenta, “The Evolution of Medical Opinion on the Use of Emetics,” press release, July 1, 2023, \url{https://www.syngenta.com/sites/default/files/The\_evolution\_of\_medical\_opinion\_on\_the\_use\_of\_emetics.pdf}.}}
\end{figure}

Deaths by inhalation or ingestion, suicidal or otherwise, are often quick and virtually unstoppable; there is still today no antidote. But their speed has masked a chemical reaction much deeper in the brain, one that ICI’s successor, Syngenta AG, would engage the scholarly community to deny. ICI knew as early as 1958 that paraquat kills weeds for the same reason it can kill cells in the mammalian brain: in both cases redox cycling triggers catastrophic oxidative stress, and ICI’s own scientists were aware of such neurological harms.\endnote{J. C. Gage (Medical Department, ICI) to F. F. Snowdon (Technical Department, ICI), October 13, 1958, 1, Bates no. SYNG-PQ-23457731, UCSF IDL, \url{https://www.industrydocuments.ucsf.edu/docs/rhwc0421/}.} A semantic search asking “How did Imperial Chemical Industries (ICI) and Syngenta use science to dispute the paraquat-Parkinson’s link?” reveals the architecture of this deception.\endnote{Appendix I, “Influencing Strategy.”} In 2001, after independent researchers at the University of Rochester and the Parkinson’s Institute in California found Parkinson’s-like symptoms in mice following paraquat injection,\endnote{Philip Botham, “Video Deposition of Dr. Philip Botham in Hoffmann v. Syngenta Crop Protection, LLC, et al.,” deposition, February 25–26, 2020, 302, UCSF IDL, \url{https://www.industrydocuments.ucsf.edu/docs/znwc0421/}.} Syngenta formed a Paraquat/Parkinson’s Disease Task Team to plan what it called a “scientific influencing strategy.” This was an early version of Syngenta’s PQ SWAT Team, a rapid-response unit with legal, communications, and product safety expertise formed in 2011 to “immediately triage the situation triggered by the release of a new study or news article.”\endnote{Syngenta, “PILT PQ Communications Management,” 2011, Bates no. 502(d)-001590.0004, UCSF IDL, 1, \url{https://www.industrydocuments.ucsf.edu/docs/xpwc0421/}; for “scientific influencing strategy (ongoing)” see Bates no. SYNG-PQ-00481740, cited in Michaels, Expert Report, PDF p. 105.}

INDRA’s synthesis is instructive here because when asked to perform a semantic search to reveal scholarly conflicts of interest, it combines its analysis of the industry documents with a database and tracker we call Transparency, a toolkit that queries OpenAlex, PubMed Central (via NCBI’s EFetch API), and other publicly accessible databases of journal article metadata. Transparency harvests competing interest disclosures wherever journals have published them, and combines these with citational information from Elsevier’s Scopus API and publicly available news sources—with INDRA’s watchdogs in tow.\endnote{Transparency uses an internal corpus, not an external RAG. Disclosures harvested from OpenAlex and PubMed Central are combined with citational data from Scopus, and then vetted for funder and author identity. Authors are anchored to ORCID, and the dataset is held locally after vetting. No outbound requests are made at analysis time, and the LLM has no ability to access external networks during analysis sessions. This preserves the integrity of the sandbox by ensuring that nothing is fetched mid-session. The database is consulted only when a question calls for it, or when a user invokes the bio: operator in Boolean search. See also Conrad Safranek et al., “ChatGPT for Automated Cross-Checking of Authors’ Conflicts of Interest Against Industry Payments,” \emph{Otolaryngology–Head and Neck Surgery} 170, no. 6 (2024): 1512–18. Safranek et al. used an LLM to compare disclosures for 78 articles against five firms’ Open Payments records. Transparency, by contrast, maintains a persistent, corpus-wide graph in which every tie is anchored to its disclosing text, verified deterministically, and quarantined from aggregates when verification fails.} The result is a knowledge graph and searchable map documenting the financial and scholarly networks underwriting global regulatory neglect. This tool, combined with a new search operator, “bio:”, enables the near-instant collection of data concerning researchers’ potential conflicts of interest. When asked, “Which studies evaluate the paraquat-Parkinson’s link, what does each conclude, and what relationships did their authors have to herbicide manufacturers?” INDRA consults its Transparency database, which provides us with a list of industry-funded scientific papers on the subject, none of which finds any solid evidence for a link between paraquat and Parkinson’s. A good example is the study published by Li et al. in 2005, written by scientists at the product-defense firm Exponent, Inc., funded by CropLife America, and thanking Syngenta’s Charles Breckenridge for “helpful discussions.” INDRA reports that in emails, Li offered with regard to paraquat, “it may be possible to weaken its direct association with PD.”\endnote{Abby A. Li, Pamela J. Mink, Laura J. McIntosh, M. Jane Teta, and Brent Finley, “Evaluation of Epidemiologic and Animal Data Associating Pesticides with Parkinson’s Disease,” \emph{Journal of Occupational and Environmental Medicine} 47, no. 10 (2005): 1059–87. For “weaken its direct association,” see Bates no. SYNG-PQ-20791944, quoted in Michaels, Expert Report, 35. See also Syngenta Crop Protection, “Subject: Letter of Agreement,” July 16, 2009, Bates no. SYNG-PQ-01058471, INDRA Archive, \url{https://indra.stanford.edu/docs/IND-fxwp325}.}

But the pool of studies by scholars with industry ties runs much deeper. Douglas L. Weed, in a 2021 article in \emph{NeuroToxicology} with no declared conflicts, maintained that “the available evidence does not warrant a claim that paraquat causes Parkinson’s disease.” Weed in 2024 doubled down on this, finding again “no compelling scientific argument for claiming that an association exists much less a causal association.”\endnote{See (with caution) Douglas L. Weed, “Paraquat and Parkinson’s Disease: A Systematic Assessment of Recent Epidemiologic Evidence,” \emph{Medical Research Archives} 12, no. 9 (2024); Douglas L. Weed, “Does Paraquat Cause Parkinson’s Disease? A Review of Reviews,” \emph{NeuroToxicology} 86 (2021): 180–84.} But as INDRA points out, Weed in 2015 had served on Monsanto’s “Glyphosate Expert Panel,” assembled to exonerate Roundup as a cause of lymphoma. And the following year he coauthored a Monsanto-funded study in which he disclosed a litigation consultancy with the pesticide maker.\endnote{Monsanto Company, “2015 Glyphosate Expert Panel—Monsanto (IARC Roundup Response),” December 15, 2015, INDRA Archive, \url{https://indra.stanford.edu/docs/IND-ppcb392}. The panel’s review appeared as John F. Acquavella, David Garabrant, Gary Marsh, Tom Sorahan, and Douglas L. Weed, “Glyphosate Epidemiology Expert Panel Review: A Weight of Evidence Systematic Review of the Relationship between Glyphosate Exposure and Non-Hodgkin’s Lymphoma or Multiple Myeloma,” \emph{Critical Reviews in Toxicology} 46, sup1 (2016): 28–43, \url{https://doi.org/10.1080/10408444.2016.1214681}; see also the 2018 corrigendum, \url{https://doi.org/10.1080/10408444.2018.1522142}, disclosing two years late that Monsanto’s own William Heydens reviewed the initial draft, a contribution omitted from the original version. Even before Weed served on the Monsanto panel, his private consultancy, DLW Consulting Services, LLC, participated in the Health and Environmental Sciences Institute’s subcommittee on Evaluating Causality in Epidemiologic Studies in 2012–2013, for which participating organizations included Monsanto Company and Syngenta Crop Protection, Inc., alongside Dow Chemical, ExxonMobil Biomedical Sciences, Bayer CropScience, Shell Oil, and others. HESI, “Evaluating Causality in Epidemiologic Studies Subcommittee,” accessed August 1, 2026, \url{https://hesiglobal.org/wp-content/uploads/2020/08/EPIDEMIOLOGY-SUBCOMMITTEE.pdf}.} Alongside him on that panel was Sir Colin Berry, whose 2010 paper “Paraquat and Parkinson’s Disease” cast doubt on any solid evidence for a paraquat-PD link.\endnote{\enlargethispage{2\baselineskip}See (with caution) Colin Berry, Carlo La Vecchia, and Pierluigi Nicotera, “Paraquat and Parkinson’s Disease,” \emph{Cell Death \& Differentiation} 17, no. 7 (2010): 1115–25, \url{https://doi.org/10.1038/cdd.2009.217}. For Berry’s membership on the Glyphosate Expert Panel, see Gary M. Williams et al., “A Review of the Carcinogenic Potential of Glyphosate by Four Independent Expert Panels and Comparison to the IARC Assessment,” \emph{Critical Reviews in Toxicology} 46, sup1 (2016): 3–20.} INDRA points to a confidential slide deck from 2009, in which Syngenta states that “Last year, Syngenta consulted with several independent senior medical and epidemiological scientists under the guidance of Professor Sir Colin Berry [and] Professor Pierluigi Nicotera… Their conclusion was that no such link existed and a publication summarizing these conclusions will be issued in 2010”\endnote{Jonathan Sullivan, Lewis Smith, and Gerardo Ramos, “Paraquat Update—Syngenta Executive Committee Meeting Nov. 9, 2009,” November 9, 2009, UCSF IDL, Bates no. SYNG-PQ-13131087, \url{https://www.industrydocuments.ucsf.edu/docs/zmwc0421/}; Carey Gillam, “Revealed: The Secret Push to Bury a Weedkiller’s Link to Parkinson’s Disease,” \emph{The Guardian}, June 2, 2023, \url{https://www.theguardian.com/us-news/2023/jun/02/paraquat-parkinsons-disease-research-syngenta-weedkiller}.} (see Figure 6). Berry, who failed to disclose this work, or his 2002 work for Syngenta on its Extended Health Science Team, has elsewhere dismissed concern over carcinogenic chemicals in colorful terms: “It’s a bit like the search for witches. You can always find them.”\endnote{Dick Thompson, “The Danger in Doomsaying,” \emph{Time}, March 9, 1992, UCSF IDL, \url{https://www.industrydocuments.ucsf.edu/docs/zqnf0116/}. A decade later Berry, by then former chairman of the U.K. Advisory Committee on Pesticides, wrote the foreword to \emph{Gaining Consumer Confidence: Residues of Crop Protection Products in Food}, a booklet Syngenta Crop Protection Ltd. distributed to growers and retailers in July 2002; cited in Kathy Hammond, “Gaining Consumer Confidence,” Fresh Produce Journal, July 22, 2002, \url{https://www.fruitnet.com/fresh-produce-journal/gaining-consumer-confidence/102326.article}.}

Using Boolean, Ecology, and semantic searches together with the Transparency tracker can shed light on how old poisons come to saturate new pastures. Paraquat litigation is ongoing, with new documents awaiting discovery, and we are only just beginning to explore the advantages of bringing these disparate datasets into conversation with one another. Industry-funded scholarship, much like litigation testimony, can be lucrative and is largely invisible: consulting and testimony are not indexed on Google Scholar and are often omitted from publicly available websites or CVs—or may even be subject to confidentiality agreements. Combining the techniques used here, we can begin to form an honest accounting of this shadow circuit.

\begin{figure}[H]
\centering
\fbox{\includegraphics[width=1.00\textwidth,height=0.72\textheight,keepaspectratio]{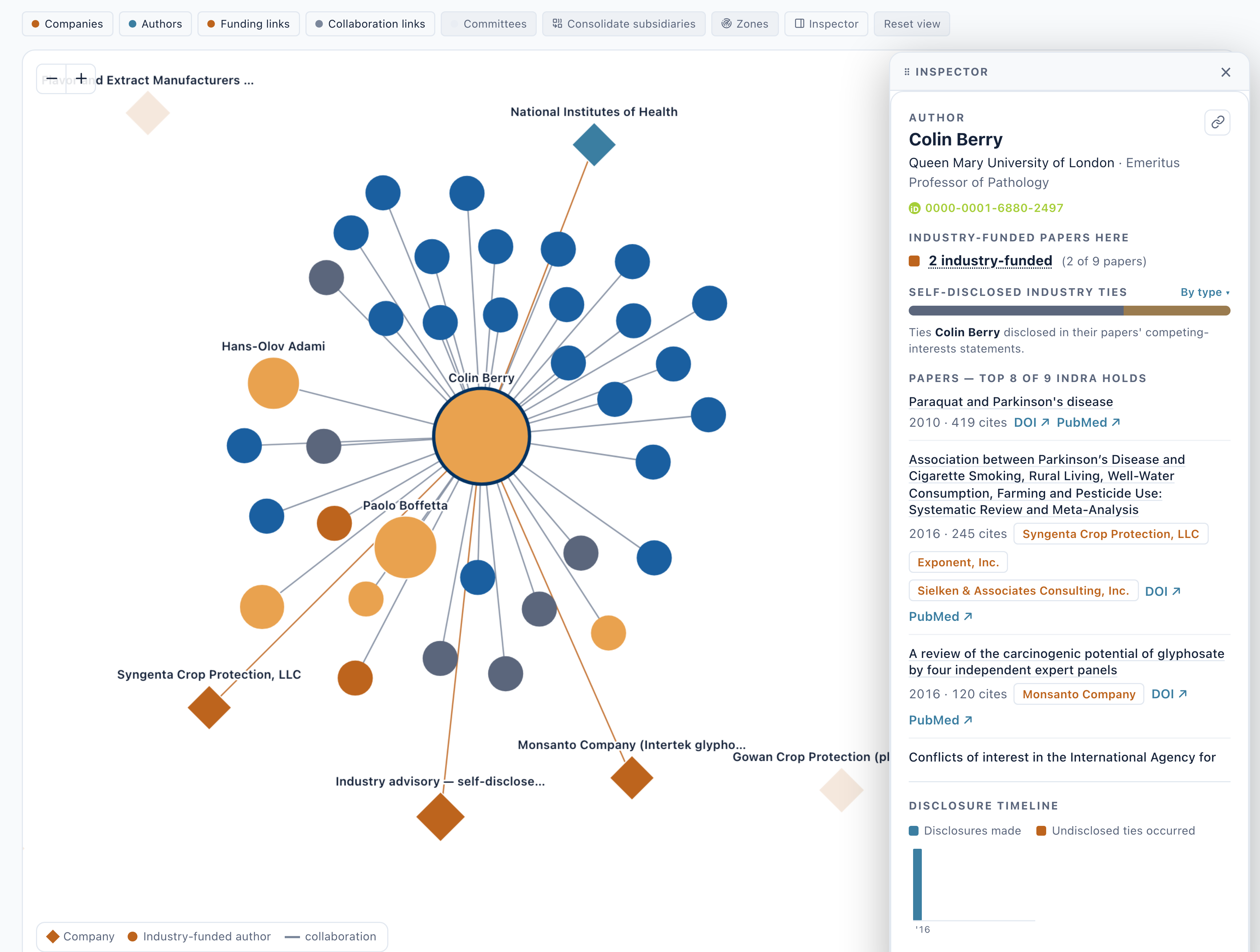}}
\caption{The industry ties of Sir Colin Berry. INDRA’s Transparency tool combines publicly available journal article metadata with industry documents, making it possible to automate the identification and verification of industry funding. When INDRA quotes from a peer-reviewed study that discloses industry funding, it marks this with an in-line [T] tag for Transparency, clarifying that there is a potential conflict of interest.}
\end{figure}

\FloatBarrier
\section{Discussion}

LLMs are making new kinds of scholarship possible—and new kinds of scholarly abuse. Recent years have seen a deluge of journalism, scholarly articles, and even entire PhD theses written by or processed through AI, for better or for worse. At the same time, it is important to appreciate how LLM-based tools can facilitate the interrogation of archives in ways previously not possible—and generate novel findings of significance.

New, for example, is the ability to de-paternalize our engagement with massive online archives, and to decouple such collections from the methods used in traditional keyword search. In a pre-LLM world, researchers had to rely on finding aids furnished by the archive’s designers, with the only route to any given set of documents being the corridors (physical and digital) made available by curators and in a very specific and heavily gated manner. In an old-style brick-and-mortar library, the researcher would look up a term or topic in the card catalog and then descend into the stacks to find whatever the index held in store. Of course, there was often a fortunate serendipity to this: accidental discoveries could be made by random wanderings—albeit only with permission from the holding institution. The creation of publicly accessible online archives in the 1990s meant that, for the first time, one could search millions of pages of text for any term regardless of whether it had ever been included in an index or catalog. Researchers gained the ability to search for and find words or phrases quite apart from those imagined by cataloging archivists. These new online archives meant that rhetorical needles could now be pulled from near-infinite haystacks.\endnote{Robert N. Proctor, “God Is Watching: History in the Age of Near-Infinite Digital Archives,” \emph{Journal of Public Health Policy} 39, no. 1 (2018): 24–26.}

INDRA builds on this capacity by making it possible for researchers to take such threads and weave them into an interpretive (and narrative) fabric, evidence-based and verifiable. Multiple archives can now be placed in conversation with one another in ways that enable the underlying LLM to generate interpretations useful to the researcher. As we have seen, tools of this sort can help draw out new rhetorical patterns, modes of concealment, and networks of collaboration from documents never intended to leave the black box of some corporate headquarters. The walls between previously siloed collections are also starting to collapse as entire archives—from the manufacturers of plastics, pesticides, food, firearms, and drugs—can now be sifted, sorted, summarized, and read against one another. Building bridges between archival collections of this sort can work to the benefit of all of us dedicated to scholarly integrity.

The archives engaged here are ever expanding, along with the kinds of questions we can ask. Whom does Chevron consider to be its closest political allies? How much money did Philip Morris give to the Whitney Museum of American Art and for what purpose? Why did the head of the Sugar Research Foundation help organize the cigarette conspiracy, and why did the Tobacco Institute say that smokers needed “anti-lynch protection”? 

The kinds of commands we can give are equally varied. One can ask for all examples of exaggeration, nostalgia, fear, anger, and affection expressed in the minutes of BP’s board meetings from 1998 to 2000. Or for a typology of the rhetorical strategies used by Exxon to disparage reports of the IPCC. We can summon an army of agents to ask: how does Meta use “edutainment,” “open loops,” and “infinite scroll” to keep users addicted to Facebook? And how do the makers of large language models hide the environmental costs of hyperscale data centers like xAI’s Colossus I or OpenAI’s Project Jupiter? Any of these and countless other questions can be asked or answered in dozens of different languages, with the results displayed in any desired form: a Shakespearean sonnet or the language of a four-year-old, or as “load-bearing” neuralese (i.e., clankerspeak), etc.

And where is this all going? As of August 2026, publicly available large language models of any quality are limited to analyzing one million tokens at a time, but future models with expanded capacity could dramatically change that.\endnote{Meta’s open-weight model, Llama 4 Scout, today advertises a ten-million-token window, but recall degrades badly as the context window fills, making it unsuitable for archival research.} We should also note that nothing in our design depends on any one LLM: the same sandbox, provenance, and verification layers can be built over open-weight models and run locally. Open-source, open-weight models may well be preferable for institutions wanting to take on sensitive investigations bound by confidentiality, and may even allow us to peer under the hood to better understand how a model derives its conclusions.

This ground is shifting rapidly. It is already possible to identify visual elements in a large set of cigarette ads (birds, bikinis, or barbecues on the beach) and soon we should be able to identify and geolocate all of the letters or memos typed using a specific typewriter, say, at Dow Chemical’s Midland headquarters or Union Carbide in Manhattan. Eventually, we should be able to analyze trial video testimony and identify all instances of laughter, anger, frustration, or even dissimulation by expert witnesses or corporate reps. Tools of this sort are a bit like the world’s first telescopes and microscopes: those who first built them had little idea of what worlds might be revealed.

\FloatBarrier
\section{Limitations}

The most important limitation we encounter when interrogating massive online archives is the maximum number of tokens that can be analyzed at any one time, a limitation of all current LLMs. This requires the researcher to select a body of documents prior to compiling them for analysis. Because only 500,000 tokens can be reliably analyzed at a time—roughly equivalent to five scholarly books—the researcher must select a subset from the archive. A researcher who searches “‘Remington’ AND ‘right to bear arms’” has already made a decision that determines which documents will enter the sandbox; different search terms produce different corpora, and different corpora will yield different results.\endnote{Lara Putnam, “The Transnational and the Text-Searchable: Digitized Sources and the Shadows They Cast,” \emph{The American Historical Review} 121, no. 2 (2016): 377–402.} We expect this token-limit constraint to become less important as AI providers allow ever-larger numbers of tokens to be analyzed. But for now, LLMs force a discipline of source selection that mirrors the historian’s practice of collecting and analyzing portions of archives that might, altogether, span thousands of linear feet.\endnote{In the current design, researchers who Bring Their Own API Key (BYOK) can access 550,000 tokens of context in any given session (500,000 tokens of space in the context window is reserved for documents). Researchers using the house key (Claude Haiku 4.5) are able to access 130,000 tokens at a time (100,000 tokens of space in the context window is reserved for documents), closer to the size of a single book.}

Another limitation (bonus?) is what we call the \textbf{Heraclitus effect}—the fact that an LLM will rarely answer a complex question exactly the same way twice. Even with the same question posed to the same set of documents, an LLM will often provide slightly different responses because the math that enables its next-word prediction is spread across thousands of processors that do not finish in the same order twice. LLM output generation is fundamentally non-deterministic, meaning that even identical queries can emphasize different passages from the selected documents. INDRA mitigates this with grounded generation, verbatim extraction, and watchdog scripts (see Methods). Good research practice always involves testing and retesting using different methods and datasets, much as the carcinogenicity of cigarettes was established by multiple and converging lines of evidence. We may never set foot in the same river twice, but that is not necessarily a bad thing. A river being different each time is not a defect; it’s what rivers do.

Related to the Heraclitus effect is what we call the \textbf{steppingstone dilemma}, a form of path dependency inherent in conversational AI. Within any given Q\&A session, early prompts establish a framing, vocabulary, and set of assumptions that shape the model’s subsequent behavior. If a user begins with ideologically loaded framing, the model treats that as a signal of what kind of conversation is underway and adjusts accordingly. This is related to the well-documented phenomenon of LLM sycophancy, the tendency of models to tell users what they think they want to hear. We call this the steppingstone dilemma because if you are trying to cross a river dotted with steppingstones, the steps you take will determine which stones you will be able to reach next. This is much like the well-known bias involved in any public opinion survey that ignores question-order effects—because people learn from the questions asked early on in the survey.\endnote{Howard Schuman and Stanley Presser, \emph{Questions and Answers in Attitude Surveys: Experiments on Question Form, Wording, and Context} (1981; repr., Thousand Oaks, CA: Sage, 1996).} A user who takes their first steps in a particular direction has already shaped the model’s subsequent path—and aptitude (Laban et al. note that LLMs “get lost in multi-turn conversation”).\endnote{Philippe Laban, Hiroaki Hayashi, Yingbo Zhou, and Jennifer Neville, “LLMs Get Lost in Multi-Turn Conversation,” in \emph{Proceedings of the International Conference on Learning Representations} (ICLR, 2026), 54738–78.} Thankfully, an LLM session costs very little to restart: INDRA caps the number of queries per session and restricts session duration so that the user can exit and start fresh again and again, allowing a kind of virtuous forgetting on the part of the platform.

A further limitation is the \textbf{gullibility (or mafia) problem}, which stems from the AI’s willingness to believe dishonest or misleading texts. If you ask INDRA what is meant by “it would be a shame if something were to happen to Jonny’s car,” it correctly identifies this as a veiled threat, but there are other situations where the AI is less perceptive. When cigarette makers assert privately that “smoking is an adult custom only,” INDRA may take this at face value rather than recognizing it as a calculated and dishonest performance of concern (i.e., eavescasting). Cigarette makers by the 1970s were starting to realize that their documents might one day be made public, which is why they mandated the use of guarded language in memos and internal reports, especially after the Federal Rules of Civil Procedure were amended (in 1970), expanding the scope of discovery in litigation.\endnote{Brown \& Williamson, “Records Management Programme: Records Creation,” n.d., 15, UCSF IDL, \url{https://www.industrydocuments.ucsf.edu/docs/fkkv0035/}; British American Tobacco, “Record Management Programme: Training Seminar, Geneva, Switzerland,” June 30–July 5, 1991, UCSF IDL, \url{https://www.industrydocuments.ucsf.edu/docs/hydb0213/}.} The gullibility problem has a counterpart in the legal fiction that documents “speak for themselves”—a common courtroom instruction that presumes transparency of intent and honesty within the four corners of any given document. INDRA, like an inexperienced judge or juror, may not always recognize the difference between what the industry says and what it actually intends. When provided with snippets from more than a hundred issues of Exxon employee magazines, the model may take viewpoints expressed in the articles too literally, at face value. In one test session, INDRA claimed that “climate” was considered by Exxon to be “an actively debated but taken-seriously physical phenomenon, developed through Exxon Research \& Engineering’s own modeling work.” 

We know from decades of scholarship, however, that by the 1980s Exxon had launched a campaign to undermine the scientific consensus regarding anthropogenic climate change.\endnote{Oreskes and Conway, \emph{Merchants of Doubt}; Geoffrey Supran, Stefan Rahmstorf, and Naomi Oreskes, “Assessing ExxonMobil’s Global Warming Projections,” \emph{Science} 379, no. 6628 (2023), \url{https://doi.org/10.1126/science.abk0063}.} A closed-context LLM with its context window packed with industry propaganda can easily fall victim to a naïve misunderstanding, especially when prompted to remain unbiased. This is particularly important when dealing with adversarial archives, where documents are not necessarily honest or innocent. Historians are familiar with “archival silence” and the inevitability of collection bias,\endnote{Matthew Connelly, \emph{The Declassification Engine: What History Reveals about America’s Top Secrets} (New York: Pantheon, 2023).} but deliberate eavescasting adds an additional layer of underhandedness that a stochastic parrot may not comprehend or appreciate.

One final limitation derives from the “conventional wisdom” of crowds on the open internet, a phenomenon AI researchers call \textbf{domain knowledge contamination}. Even within the evidentiary sandbox, the model cannot be entirely detached from epistemologies or ideologies it learns (at training time) by scraping the open internet. It must rely on what it already knows about the meanings of terms, and we cannot expect it to disregard popular knowledge about the Marlboro Man or the fact that “Winston tastes good like a cigarette should.” INDRA’s [DK] and [INT] markers are designed to make such leakage visible, but they cannot eliminate it. Just as jurors in a tobacco trial are asked to set aside what they already know and rely purely on evidence presented in court—a legal fiction, since no one can shed a lifetime of cultural experience—INDRA’s evidentiary sandbox is a powerful constraint but not an airtight seal. Keep in mind also that there is both virtue and vice in conventional wisdom, and that domain knowledge is generally indispensable; indeed, there could be no LLM reasoning without it. Looking at only one tiny slice of the library, for example, INDRA might conclude that some event or rhetorical form occurred first in 1994, but that might be only because it has been fed documents from a certain date range. In such cases, the model’s prior knowledge that, say, smoking was allowed on American nuclear submarines until 2011 (but never on French or German subs) can help establish an honest chronology.\endnote{Tim Henderson and Robert N. Proctor, “Why Was Smoking Allowed on U.S. Submarines for More Than a Century?” (manuscript under review).}

Of course, humans also suffer most or even all of these same limitations: selection bias, gullibility, conversational path dependency, sycophancy, information overload, “lost in the middle” neglect, and so forth. Conventional wisdom is often unreliable, and humans don’t necessarily tell the same story in exactly the same way twice. And hopefully never will.

\FloatBarrier
\section{Conclusion}

{\looseness=-1 Etymology reminds us that there is paradise in a walled garden (\emph{paridaida }in Old Persian), and searching with INDRA is a bit like foraging in such a garden. The methods developed here represent a new kind of close reading, one could say \emph{closed reading}, in which the researcher gains interpretive power and speed by accepting creative constraints.\endnote{Ilse Aichinger, “The Bound Man” (“Der Gefesselte”), in \emph{The Bound Man and Other Stories}, trans. Eric Mosbacher (London: Secker \& Warburg, 1955).} What emerges is a structured interlocutor for archival corpora, built on the principle that with proper grounding, an LLM can become a reliable research partner rather than a mere chatbot or stochastic confabulator. \par}

One contribution here is to decouple the documents contained in massive online archives from the gated platforms that have until now dictated how they must be accessed and searched. A further contribution lies in the deployment of an LLM with safeguards heretofore absent from archival research: an evidentiary sandbox, an in-line provenance-tagging convention, watchdog scripts, and a transparency tracker capable of disclosing potential conflicts of interest. Our three case studies show what kinds of things can be uncovered: a “college” sworn to secrecy, a Big Carbon newsletter denying climate science, a network of scholars paid to exonerate a deadly poison—and the silences that so often reveal evidence of corporate malfeasance.

{\looseness=-1 Adversarial archives raise the stakes of such inquiries. Documents produced by connivance or court order cannot always be taken at face value, and any tool built to read them must consider carefully how and why they were created, and how and why they came to light. On the one hand, an LLM whose context window is loaded up with industry talking points might draw misleading inferences. Without proper safeguards, the result could be more quagmire than treasure trove—as in “garbage in, garbage out.” On the other hand, a promiscuous mingling with the open web can pollute an analysis in subtle ways. This is in part because we are all living in a world these industries helped create, where so many of our words (“filtered cigarette,” “carbon footprint,” “lifestyle choices”) have been coined or captured by the very actors we aim to study. In this sense, INDRA’s safeguards are best understood as a way of protecting invaluable evidence from sources of contamination.\par}

\enlargethispage{1.5\baselineskip}{\looseness=-1 Tools like this must be designed to complement, not replace, the trained scholar, and we should never hope to automate narration and interpretation in their entirety. The true context required to understand any document or set of documents is the broader social, political, and natural world that lies outside the “context window,” and that world belongs to the scholar living and breathing. The machine has no taste, no real ear for irony or humor, no eye for the visual; it lives in a soundless box, narrating its own work in a server where everything happens “silently” or “gracefully” and every detail is “smoke-tested.” The Heraclitus effect, the steppingstone dilemma, and the gullibility problem all point in the same direction: researching with AI is like working with an army of brilliant but ultimately credulous assistants, who must be watched with vigilance and occasionally corrected. As ever, the researcher must verify rather than trust—test and then test again. Purpose-built AI is powerful, but the archive still demands a reader who can recognize a sleeping dog left to lie. And who knows enough to wake it. \par}

\FloatBarrier
\section*{Acknowledgments}

Funding for this research was provided by the Climate Social Science Network at Brown University [Award \#FP-2408-02465], by the Stanford Tobacco Research Collective, and by the Grantham Foundation for the Protection of the Environment [Award \#AW1171850]. The authors wish to thank Tracey J. Woodruff, Pamela M. Ling, Robert K. Jackler, Londa Schiebinger, Lorraine Daston, Jevin West, Richard Heede, Paul Nauert, Toby Handfield, Swapneel Mehta, Haley Lepp, Logan Shaw, Arusha Patil, and the Stanford University Networks \& Organizations Workgroup for their helpful comments.

\FloatBarrier
\section*{Author Contributions}

D.A. conceived, designed, and developed the INDRA platform, wrote its software, and co-drafted the manuscript. R.N.P. contributed to the writing, analysis, and selection and development of the case studies.

\FloatBarrier
\section*{Disclosures}

INDRA is hosted by the Health \& Toxics Policy Lab at Stanford University School of Medicine, and is freely available at \url{https://indra.stanford.edu}. The platform was built using Anthropic’s Claude API; Anthropic provided no funding, API credits, technical assistance, or editorial input, and had no role in the study design. R.N.P. serves as an expert witness for plaintiffs in litigation against cigarette manufacturers.

\newpage
\begingroup
\small
\printendnotes
\endgroup
\end{document}